\documentclass{aastex631}
\usepackage{CJK}

\usepackage{tikz}

\submitjournal{ApJ}

\shorttitle{extremely low-mass white dwarfs}
\shortauthors{Wang et al.}

\begin{document}
\begin{CJK*}{UTF8}{gbsn}

\title{Extremely low-mass white dwarf stars observed in Gaia DR2 and LAMOST DR8}

\correspondingauthor{Kun Wang}
\email{kwang@cwnu.edu.cn}

\author[0000-0002-5745-827X]{Kun Wang(王坤)}
\affiliation{School of Physics and Astronomy, China West Normal University, Nanchong 637009, People's Republic of China}

\author[0000-0003-0963-0239]{P\'{e}ter~N\'{e}meth}
\affiliation{Astronomical Institute of the Czech Academy of Sciences, CZ-251\,65, Ond\v{r}ejov, Czech Republic}
\affiliation{Astroserver.org, F\H{o} t\'{e}r 1, 8533 Malomsok, Hungary}

\author[0000-0003-3736-6076]{Yangping Luo(罗杨平)}
\affiliation{School of Physics and Astronomy, China West Normal University, Nanchong 637009, People's Republic of China}

\author[0000-0001-7084-0484]{Xiaodian Chen(陈孝钿)}
\affiliation{Key Laboratory of Optical Astronomy, National Astronomical Observatories, Chinese Academy of Sciences, Beijing 100012, People's Republic of China}
\affiliation{School of Physics and Astronomy, China West Normal University, Nanchong 637009, People's Republic of China}

\author{Qingquan Jiang(蒋青权)}
\affiliation{School of Physics and Astronomy, China West Normal University, Nanchong 637009, People's Republic of China}

\author{Xingmei Cao(曹星梅)}
\affiliation{School of Physics and Astronomy, China West Normal University, Nanchong 637009, People's Republic of China}

\begin{abstract}
We present the first results from our ongoing project to study extremely low mass (ELM) white dwarfs (WDs) ($M$ $\leq$ 0.3$M_{\sun}$) 
with the Large Sky Area Multi-Object Fibre Spectroscopic Telescope (LAMOST) spectra.
Based on the LAMOST DR8 spectral database, we analyzed 136 ELM WD candidates selected from $\it Gaia$ DR2 data and 12 known objects previously identified by the ELM Survey. 
The atmospheric parameters and radial velocities of these stars were obtained by fitting the LAMOST low-resolution spectra.
After comparing the atmospheric parameters of the 12 known objects from this work to the results reported by the ELM Survey, 
we demonstrated the potential of LAMOST spectra in probing into the nature of ELM WDs.
Based on the atmospheric parameters and $\it Gaia$ EDR3 data, we identified 21 new high-probability ELM WDs 
with masses $M$ $\leq$ 0.3$M_{\sun}$ and parallax estimates that agree to within a factor of 3.
Two of them, J0338+4134 and J1129+4715, show significant radial velocity variability and are very likely to be binary systems containing at least one ELM WD.

\end{abstract}

\keywords{White dwarf stars (1799); Spectroscopy (1558); Compact Objects (288)}

\section{Introduction} \label{sec:intro}
Galactic double white dwarf (DWD) binaries, whose orbital periods are of the order of minutes to a few hours, 
are particularly interesting because they are expected to be dominant millihertz gravitational wave (GW) sources for space-based detectors,
 e.g., the TianQin Observatory \citep{Huang2020} and the Laser Interferometer Space Antenna (LISA, \cite{Amaro2017}). 
 Although millions of DWDs likely exist in our Galaxy, based on binary population synthesis studies \citep{Han1998, Lamberts2019}, 
 there are currently only about 150 DWDs with known orbital parameters \citep{Korol2021}. 
About 60\% of these systems contain an extremely low mass (ELM) white dwarf (WD) star with mass $M$ $<$ 0.25$M_{\odot}$ (see figure 1 of the work of \cite{Korol2021}).
ELM WDs are a relatively rare population of helium-core WDs ($M$ $\lesssim$ 0.3$M_{\odot}$).
They show low surface gravities of 5 $\lesssim$ $\log{g}$ $\lesssim$ 7 \citep{Li2019, Brown2020}, 
 indicating that the stars are bloated WDs and may not yet have reached the cooling track.
Compared with more massive WDs, ELM WDs have much thicker hydrogen-rich envelopes 
and may spend up to several billion years in the proto-WD phase from Roche-lobe detachment to the beginning of the cooling track \citep{Istrate2014, Istrate2016, Chen2017}.
This proto-WD phase is in the most luminous stage except for flashes.
Besides, the radii of low-mass WDs are bigger than those of canonical WDs at the same effective temperature,
based on the inverse relationship between mass and radius for WDs (e.g. \cite{Romero2019}).
 All of these keep ELM WDs brighter than their more massive counterparts and thus make them more likely to be detected by electromagnetic radiation observations \citep{Li2020}.
 
 As the age of the universe is not old enough for a single star to form such an ELM WD, 
 ELM WDs are widely believed to be the products of the evolution of close binaries \citep{Driebe1998, Althaus2013, Istrate2016, Li2019, Soethe2021}. 
 Indeed, of the currently known ELM WDs, almost 100 percent were discovered in binary systems as companions to millisecond pulsars, 
 A/F-type dwarf stars (the so-called EL CVn-type binaries, \cite{Maxted2014, Wang2018, Wang2020}), and more commonly canonical WDs. 
 The formation and evolution of ELM WDs may probably experience highly uncertain mass transfer and common envelope phases repeatedly. 
 As such, they are associated with many kinds of intriguing objects in modern astrophysics. 
 Low-mass X-ray binary systems and cataclysmic variables are potential progenitors of short-period ELM WD binaries \citep{El-Badry2021b, El-Badry2021a}. 
 About ten ELM WDs were found as companions to millisecond pulsars \citep{Mata2020}, since the first spectroscopically confirmed case reported in 1990s \citep{van1996}. 
 The fate of  ELM WD binaries depends heavily on the stability of mass transfer in the system, 
 which determines the system's merger timescale and eventual outcome \citep{Han2020}. 
 Most known double degenerate binaries with ELM WD companions are expected to merge in less than a Hubble time, 
 leading to the formation of exotic systems such as underluminous supernovae \citep{Brown2011b, Brown2016}, 
 AM CVn systems \citep{2007Deloye, Wong2021}, single hot subdwarfs of B and O type \citep{Zhang2012, Geier2022}, 
 R Corona Borealis stars \citep{Zhang2014}, and excellent sources for future space-based GW detectors \citep{Burdge2019, Burdge2020, Brown2020l}.
 As signposts of compact binary systems, a more complete sample of ELM WDs will significantly advance our knowledge of the formation channels of these very intriguing objects.
 
The search and study of ELM WDs have become one of the most active research fields in the past decade. 
\cite{Brown2010} carried out a program, called the Extremely Low Mass Survey (ELM Survey), 
aimed at spectroscopically following up ELM WD candidates selected from the Sloan Digital Sky Survey (SDSS) photometric catalog. 
The ELM Survey has identified as many as 98 DWDs, 79 of which include at least one of ELM WD component \citep{Brown2020}. 
Two ELM Survey discoveries, J0651+2844 and J0935+4411, are strong LISA sources with a signal-to-noise ratio of SNR $>$ 40 \citep{Brown2011, Brown2016, Kupfer2018}. 
By applying two different target selection techniques, \cite{Kosakowski2020} recently extended the ELM Survey into the southern hemisphere and discovered eight new ELM WDs, 
including six short-period ELM WD binary systems. Dozens of ELM WDs and their precursors were found by the WASP, PTF, and Kepler surveys (e.g. \cite{Maxted2014, van2018, Wang2019}). 
With unprecedented high-precision parallax measurements, $\it Gaia$\footnote{https://www.cosmos.esa.int/web/gaia/home} has opened a new window for ELM WD candidate selection. 
Based mainly on the second $\it Gaia$ data release ($\it Gaia$ DR2), \cite{Pelisoli2019} have created an unbiased all-sky catalogue of 5762 ELM WD candidates down to $T_{\rm eff}$ $\simeq$ 5000 K. 
However, the exact nature of these stars is still unclear and we need to obtain further spectra for this sample so as to confirm their nature as ELM WDs. 
Large multi-object spectroscopic surveys , such as the Large Sky Area Multi-Object Fiber Spectroscopic Telescope 
(LAMOST, also known as the ``Guo Shou Jing" Telescope \citep{Cui2012}), will play a very important role in this effort. 

In this work, we investigate the ELM WD candidates in the catalogue of \cite{Pelisoli2019} with LAMOST DR8 spectra. 
The atmospheric parameters of 136 ELM WD candidates are obtained by fitting LAMOST DR8 low-resolution spectra.  
After joining the results of our spectral fitting with parallax measurements from $\it Gaia$ Early Data Release 3 ($\it Gaia$ EDR3), 
we try to place constraints on the stellar nature of these objects. 

 \section{Data}\label{section2}
 \subsection{LAMOST Spectra}
The LAMOST telescope, specialized for conducting spectroscopic surveys, 
is situated at the Xinglong Station of National Astronomical Observatory, Chinese Academy of Sciences \citep{Cui2012, Zhao2012, Liu2020}. 
It can collect 4000 spectra in a single exposure in a total 20 deg$^2$ field of view.  
LAMOST data consist of two components: data from the low-resolution spectroscopic survey (LRS; $R$ $\sim$ 1800) covering the wavelength range of 3800-9100 \AA\ , 
and data from the medium-resolution spectroscopic survey (MRS; $R$ $\sim$ 7500).
For a target of the MRS, there are two spectra within an exposure, including a blue- and a red-band spectrum.
The blue and red cameras cover wavelength ranges from 4950 \AA\ to 5350 \AA\ and from 6300 \AA\ to 6800 \AA\ \citep{Liu2020, Wang2021}, respectively.
The limiting magnitude is around $\it G$ = 15 mag for targets of LAMOST-MRS \citep{Liu2020}, 
while LAMOST-LRS has a limiting magnitude of $r$ $\sim$ 18 mag \citep{Luo2015}. 
A detailed description of the LAMOST telescope can be found at http://www.lamost.org/.

\begin{figure}
\center
\includegraphics[scale=0.8]{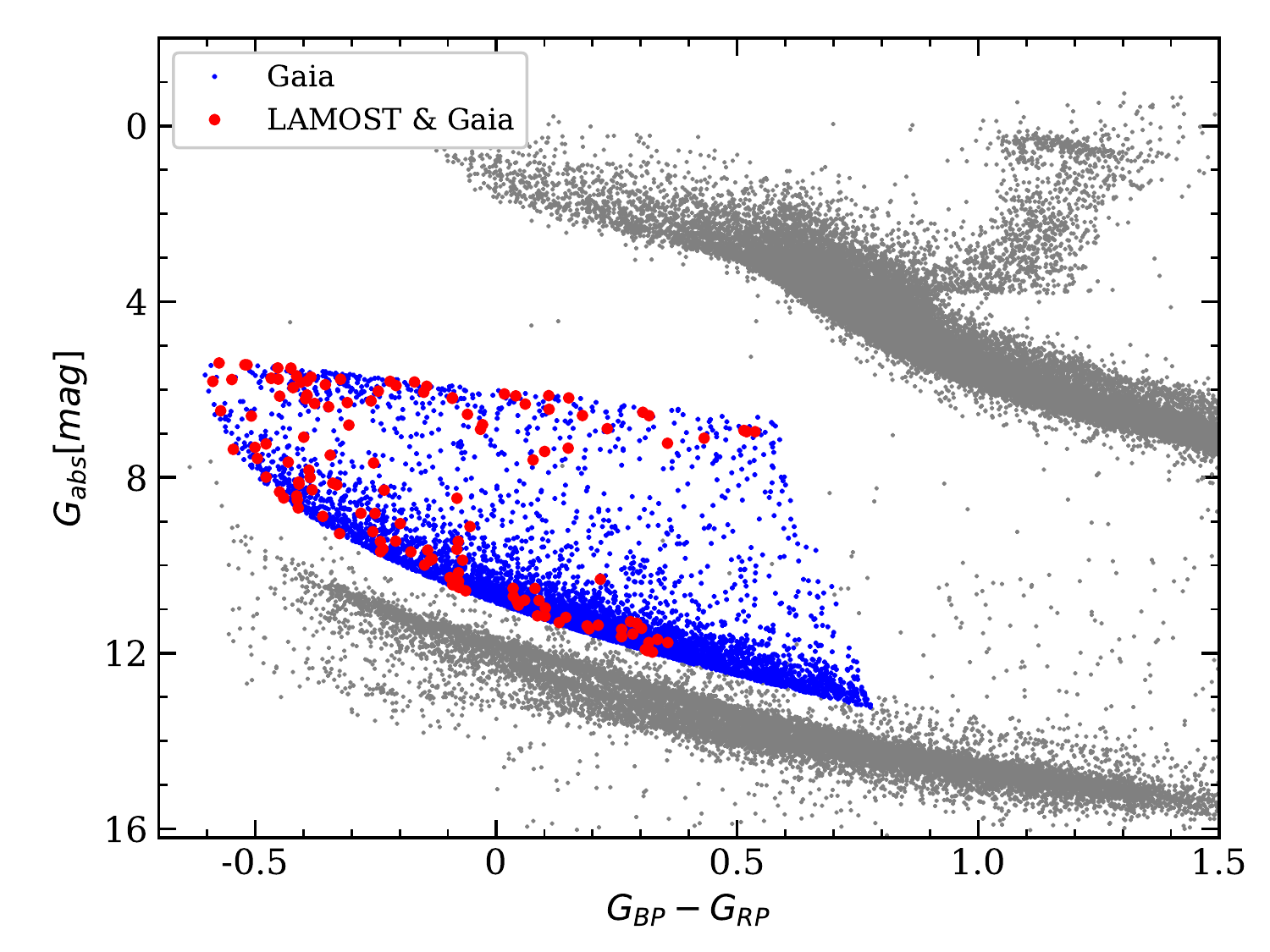}
\caption{Gaia color-magnitude diagram showing ELM WD candidates selected from the Gaia DR2. 
The blue points indicate the 5762 ELM WD candidates reported by \cite{Pelisoli2019}, whereas the red dots represent 136 of 5762 objects with LAMOST spectra.
We also show in grey the Gaia 100 pc clean sample (sample C in \citealt{Lindegren2018}) for reference, similar to \cite{Pelisoli2019}.}
\label{fig1}
\end{figure}

In 2021 March, more than 10 million spectra have been published in LAMOST DR8, including 9,396,139 stellar spectra with SNR $>$ 10. 
We cross-matched the results of \cite{Pelisoli2019} with the LAMOST DR8 Low-Resolution Survey General Catalog, 
and a total of 291 objects, including 269 ELM WD candidates and 22 known ELM WDs, were retrieved. 
We retrieved 404 low-resolution spectra for them via the LAMOST spectral archive\footnote{http://www.lamost.org/dr8/}. 
However, only 188 of the 404 spectra with a SNR higher than 10 were chosen for further analysis in order to obtain reliable results. 
In a word, we found 172 good-quality LAMOST spectra for 136 ELM WD candidates and 16 spectra for 12 known ELM WDs, respectively. 
Figure \ref{fig1} displays the positions of the 136 ELM WD candidates in a $\it Gaia$ color-magnitude diagram.
   
 \subsection{Stellar Atmosphere Fits}
 All 188 of the LAMOST low-resolution spectra were processed with the spectral fitting program {\sc XTgrid} \citep{Nemeth2012, Nemeth2019} 
 from the $\it Astroserver$\footnote{https://astroserver.org/} service. {\sc XTgrid}, a fully automated stellar spectral analysis procedure, 
 wraps around the codes {\sc Tlusty} and {\sc Synspec} \citep{Hubeny1988, Hubeny1995, Hubeny2017} to calculate non-local thermodynamic equilibrium (non-LTE) 
 model stellar atmospheres and to generate detailed synthetic spectra. 
For the ELM WDs, we employed H+He chemical compositions for all models. 
 Rather than using a pre-calculated grid of models, the default fitting procedure calculates non-LTE atmosphere models and synthetic spectra on the fly. 
 However, {\sc Tlusty} model atmospheres turned out to be inadequate below $T_{\rm eff}<14000$ K.
 For those stars, models were extracted from the ELM WD grid of synthetic spectra from \cite{Tremblay15}, or for metal rich objects, from the BOSZ spectral library \citep{bohlin17}. 
 It is worth noting here that the BOSZ library is not suitable for ELM WDs, but we cannot be sure all our stars are indeed ELM WDs.  
 The synthetic spectra were then compared to the LAMOST flux-calibrated observations with a piecewise normalization, which also cuts down systematic effects such as absolute flux inconsistencies. 
 The comparison was made by computing $\chi{2}$ values and an iterative steepest-descent $\chi{2}$ minimization method was applied for seeking for the best fitting model. 
 We fitted the 3800-6800 \AA\ spectral range, since the red part of the LAMOST spectrum contains little information for ELM WDs, but adds noise to the $\chi$$^{2}$ fit.  
 
The fitting results of the 188 LAMOST spectra, including the effective temperature $T_{\rm eff}$, surface gravity $\log{g}$, 
He abundance $y=\log{(n{\rm He}}/n{\rm H})$, and radial velocity (RV), are given in Table \ref{tab1}. 
Parameter uncertainties with 60\% confidence intervals were calculated by mapping the $\chi{2}$ variations around each best fit. 
In addition, for every LAMOST spectrum, we have listed the object's name, right ascension (RA), declination (DEC), and SNR in the $g$ band.
If an object has more than one LAMOST spectrum, all of the fitting results from the same object are given in ascending order of observation time. 
We have added notes in Table\,\ref{tab1}, such as a grade of the fit quality (A-F, in descending order), which correlates with the SNR and the spectral types of the objects. 
Spectra marked with ``PET" were fitted with ELM WD models \citep{Tremblay15}, 
those marked with ``BOSZ" were fitted with LTE models \citep{bohlin17}, and the rest were fitted with {\sc Tlusty} non-LTE models.
We marked PG1159 type objects with ``pg1159", the ones that show metal lines with ``met", nebular lines with ``neb", composite spectra with ``comp",
 and also mark stars that show C\,{\sc ii-iv} lines.
PG1159 type stars are hot H-deficient pre-WDs en route to the WD cooling sequence. 
Composite spectrum binaries, or double-lined binaries, are systems where the member stars have similar luminosities 
and contribute to the spectrum. In such systems, the members can be separated and spectroscopically investigated.
We identified composite candidates based on the presence of metal lines typical of cool companions. 
These lines are the Na I, Mg I, and Ca II lines in the optical spectrum. 
However, the mere presence of these lines in our low-resolution spectra does not necessarily indicate binarity. 
Therefore, we also considered the spectral energy distribution (SED) and the infrared excess of the source. 
Only when the SED required two components did we label it as a composite spectrum. 
The final confirmation of binarity will require high-resolution spectroscopy and RV measurements. 
Our spectroscopic data and the SED method are not suitable for resolving systems with similar components, such as DA+DA binaries.

\startlongtable
\begin{deluxetable*}{lrrrrrrrr}
\tablecolumns{8}
\tablecaption{Measured parameters for the 188 LAMOST low-resolution spectra\label{tab1}}
\tabletypesize{\scriptsize}
\setlength{\tabcolsep}{2.5pt}
\tablewidth{0pc}
\tablehead{\colhead{Object} & \colhead{RA$_{J2000}$} & \colhead{DEC$_{J2000}$} & \colhead{$T_{\rm eff}$} & 
\colhead{$\log{g}$} & \colhead{$y^{a}$} & \colhead{RV} & \colhead{SNR$_{g}$} & \colhead{Notes}\\
\colhead{}        &\colhead{(degree)}      &\colhead{(degree)}        &\colhead{($\rm K$)}     &\colhead{($\rm cm~s^{-2}$)}    &   &\colhead{($\rm km~s^{-1}$)}  &\colhead{}  &\colhead{} 
} 
\startdata
J0000+4653      	& 0.222364 & 46.893559 		& 37900$\pm$2420 		& 5.966$\pm$0.296 		& 0.903$\pm$0.214  		& -115.8$\pm$14.8    &18.9 &C\\
J0000+4653   		& 0.222364 & 46.893559 		& 34500$\pm$1020 		& 5.426$\pm$0.198 		& 1.020$\pm$0.002    	& -128.6$\pm$10.0   & 20.1 &B\\
J0008+5039     		& 2.202916 & 50.663614 		& 114540$\pm$31380 	& 8.228$\pm$0.824 		& -2.300$\pm$0.281  	& -448.5$\pm$15.6   &19.4 &C\\
J0008+5039    		& 2.202916 & 50.663614 		& 60000$\pm$11000   	& 8.000$\pm$0.500   	& -1.881$\pm$0.056   	& 329.0$\pm$150.0   	& 13.6 &D\\
J0012+2747		& 3.135519 & 27.785974 		& 23430$\pm$200    		& 7.270$\pm$0.042   	&$<$ -3.768     			&4.4$\pm$6.9   	&37.6 &A\\
J0019+5105 		& 4.975141 & 51.084907    	& 64600$\pm$9660 		& 5.649$\pm$0.210     	& -1.282$\pm$0.282   	& -8.4$\pm$11.5    	& 12.2 &C\\
J0026+4433		& 6.528333 & 44.555064    	& 36130$\pm$500   		& 5.673$\pm$0.104   	& -1.638$\pm$0.051   	& -39.1$\pm$4.1       & 26.9 &A\\
J0043+2850 		& 10.865785 & 28.846921  	& 18310$\pm$460   		& 6.662$\pm$0.098      	& -1.907$\pm$0.498  	& 19.5$\pm$7.8        & 19.0 &A\\
J0045+2121 		& 11.296490 & 21.352159    	& 72740$\pm$3420     	& 7.044$\pm$0.201      	& $<$ -5.503    			& -65.2$\pm$30.0     & 10.2 &C\\
J0046+3433 		& 11.617971 & 34.555527  	& 15030$\pm$320         	& 7.263$\pm$0.049      	& -1.153$\pm$1.135    	& -16.2$\pm$7.0       & 17.1 &A\\
J0054+3750		& 13.666097 & 37.843652        & 112000$\pm$8000    	& 8.500$\pm$0.500        	& -2.401$\pm$0.421         & 47.9$\pm$25.5       & 23.8 &D\\
J0054+3750 		& 13.666075 & 37.843638        & 62000$\pm$10000     	& 8.000$\pm$0.500     	&$<$ -2.129    			& 45.1$\pm$30.0      & 10.7 &D\\
J0058+4454		& 14.654559 & 44.908567        & 15440$\pm$900       	& 5.594$\pm$0.148        	&$<$ -2.532     			& 195.7$\pm$7.5      & 23.5 &A\\
J0101+0401	 	& 15.369537 & 4.033043          & 11380$\pm$70       		& 5.226$\pm$0.098      	& -4.825$\pm$0.191         & -131.7$\pm$5.4   	& 18.9&PET,A\\
J0109+5249 		& 17.418584 & 52.819013        & 34660$\pm$310        	& 5.943$\pm$0.065       	& -1.412$\pm$0.133         & -25.4$\pm$7.8       & 31.7 &A\\
J0211+4659 		& 32.999152 & 46.993375        & 70600$\pm$2120      	& 6.380$\pm$0.078         	& -2.382$\pm$0.159         & -87.7$\pm$2.2       & 24.8 &A\\
J0215+0155 		& 33.776147 & 1.917815          & 10540$\pm$40         	& 5.055$\pm$0.067           &$<$ -4.635     			& -65.4$\pm$9.2       & 17.5 &PET,A\\
J0228+1915 		& 37.037323 & 19.252703    	& 46590$\pm$1070       	& 7.580$\pm$0.106       	&$<$ -4.078      		& -9.5$\pm$6.4         & 34.7 &A\\
J0230+2922 		& 37.530274 & 29.382299        & 85460$\pm$3360        	& 8.796$\pm$0.273     	& -0.516$\pm$0.068          & -24.9$\pm$17.5    & 36.2 & B\\
J0238+4123 		& 39.600227 & 41.388995        & 80210$\pm$12440      	& 6.246$\pm$0.333     	&$<$ -3.396        		& -129.1$\pm$22.5   & 15.8 &B\\
J0238+4123 		& 39.600227 & 41.388995        & 95010$\pm$16830      	& 6.393$\pm$0.238     	&$<$ -4.128      		& 112.8$\pm$27.0    & 10.2 &C\\
J0254+4750 		& 43.517385 & 47.835969        & 33050$\pm$160          	& 7.008$\pm$0.028     	& -3.624$\pm$0.108          & -15.3$\pm$5.2      & 60.1 &A\\
J0308+5140 		& 47.075785 & 51.669867        & 7650$\pm$100          	&$>$ 5.000     			&-0.911$\pm$0.079           & 25.0$\pm$1.9         & 13.9 &BOSZ,met,B\\
J0312+4551		& 48.085425 & 45.856474        & 18610$\pm$310          	& 7.107$\pm$0.074     	&$<$ -3.092       		& -53.8$\pm$18.3      & 18.7 &A\\
J0313+4131 		& 48.294554 & 41.522199        & 16050$\pm$430           	& 7.377$\pm$0.095     	& -2.018$\pm$0.389      	& 29.1$\pm$6.0     & 11.9 &A\\
J0314+4812 		& 48.691328 & 48.201636        & 97390$\pm$9070         	& 6.732$\pm$0.282     	&$<$ -5.689 			& 92.7$\pm$18.9     & 19.2 &C\\
J0318$-$0107 		& 49.555241 & -1.119923         & 15680$\pm$300           	& 7.756$\pm$0.087     	&$<$ -2.631       	        & 82.3$\pm$3.2       & 97.8 &A\\
J0318$-$0107 		& 49.555319 & -1.119922         & 15010$\pm$160               & 7.877$\pm$0.047     	& -4.857$\pm$0.229         &84.8$\pm$5.1         & 65.1 &PET,A\\
J0318$-$0107		& 49.555491 & -1.119941      & 13480$\pm$150            	& 7.902$\pm$0.135     	& -4.962$\pm$0.330         & 72.0$\pm$9.4       & 56.0 &PET,A\\
J0338+4134 		& 54.696112 & 41.573393      & 21920$\pm$300            	& 5.723$\pm$0.042     	&$<$ -4.065       		& -104.4$\pm$3.1   & 80.5 &A\\
J0338+4134		& 54.696112 & 41.573393      & 22060$\pm$260            	& 5.792$\pm$0.037     	&$<$ -3.446       		& 79.4$\pm$3.3        & 67.8 &A\\
J0341+4938		& 55.439619 & 49.644315     & 33160$\pm$770            	& 7.408$\pm$0.165     	& -2.321$\pm$0.074         & 121.2$\pm$13.6  & 13.6 &B\\
J0344+2514		& 56.092744 & 25.248134    & 28340$\pm$500             	& 6.287$\pm$0.097     	& -2.646$\pm$0.038         & -41.6$\pm$6.1     & 21.0 &A\\
J0344+2514		& 56.092744 & 25.248134    & 27390$\pm$560             	& 6.111$\pm$0.108      	& -2.388$\pm$0.556         & -6.1$\pm$6.7       & 13.8 &A\\
J0346+1555		 & 56.621058 & 15.926522    & 31760$\pm$290             	& 7.574$\pm$0.047     	& -3.128$\pm$0.161         & 23.1$\pm$8.4       & 20.4 &A\\
J0349$-$0059		 & 57.322536 & -0.988681    & 95000$\pm$5000           	& 8.000$\pm$0.500    	& -0.576$\pm$0.200         & -88.6$\pm$16.9    & 20.8 &pg1159,D\\
J0354+3515		& 58.700920 & 35.255844      & 72400$\pm$4630           	& 6.670$\pm$0.144        	& -1.469$\pm$0.131          & 28.0$\pm$13.8     & 39.0 &B\\
J0354+3515		& 58.700920 & 35.255844      & 74340$\pm$4090           	& 5.812$\pm$0.135      	& -1.867$\pm$0.160            & 29.6$\pm$17.2     & 13.4 &C\\
J0404+2917		& 61.015545 & 29.284396    & 25020$\pm$690             	& 7.725$\pm$0.095      	&$<$ -2.176       		& 51.6$\pm$14.1     & 19.9 &A \\
J0404+0800		& 61.076658 & 8.000846      & 10860$\pm$120              	& 4.985$\pm$0.121     	&$<$ -4.634                       & 51.6$\pm$6.9         & 30.3 &PET,met,A\\
J0414$-$0240		 & 63.513183 & -2.676125    & 15950$\pm$460              	& 7.297$\pm$0.117      	&$<$ -1.486        	          & -18.6$\pm$12.6   & 10.5 &A\\
J0426+4820		& 66.598462 & 48.335746   & 46840$\pm$670              	& 8.148$\pm$0.091      	& -1.964$\pm$0.009           & 45.8$\pm$22.9    & 13.1 &B\\
J0427$-$0253		 & 66.750345 & -2.883789    & 25230$\pm$1370            	& 5.584$\pm$0.148      	&$<$ -3.558         		  & 13.7$\pm$12.3    & 16.2 &B\\
J0431+5538		& 67.871440 & 55.642241      & 39900$\pm$1320            	& 7.771$\pm$0.272      	&$<$ -2.309         		   & -52.7$\pm$19.2     & 12.6 &C\\
J0431+5538		& 67.871288 & 55.642414    & 42670$\pm$1150            	& 7.791$\pm$0.096      	&$<$ -2.645         	           & 21.3$\pm$16.3      & 37.8 &A\\
J0439+3954		& 69.750106 & 39.901196    & 18040$\pm$710              	& 6.712$\pm$0.284      	& -2.464$\pm$1.197            & -20.9$\pm$15.0     & 10.6 &B\\
J0442+4111		& 70.632603 & 41.186821    & 69890$\pm$6080            	& 6.275$\pm$0.209      	& -1.847$\pm$0.258             & -89.5$\pm$13.0    & 14.1 &C\\
J0442+4111		& 70.632586 & 41.186802    & 59740$\pm$890              	& 6.866$\pm$0.148      	& -1.320$\pm$0.131              & -49.6$\pm$9.7     & 57.6 &A\\
J0443+0541		& 70.761091 & 5.688080        & 9390$\pm$140            	& 4.998$\pm$0.097          &$<$ -4.633                           & -161.8$\pm$23.2  & 10.2 &PET,B\\
J0457+2314		& 74.366130 & 23.246784      & 25760$\pm$640              	& 7.629$\pm$0.180        	& -5.115$\pm$0.916              & 2.8$\pm$17.5       & 15.7 &B\\
J0457+2314		& 74.366295 & 23.247369    & 24570$\pm$890              	& 7.671$\pm$0.137      	&$<$ -2.738          		    & -35.7$\pm$15.6    & 14.1 &B\\
J0458+1740		& 74.732113 & 17.680364     & 16660$\pm$380              	& 6.740$\pm$0.079        	& -1.850$\pm$0.237                & 33.2$\pm$12.3       & 13.6 &B\\
J0459$-$0228		& 74.834577 & -2.468497      & 29920$\pm$710              	& 5.733$\pm$0.095      	& -2.027$\pm$0.323              & 106.9$\pm$25.9    & 10.3 &C\\
J0507+3013		& 76.992428 & 30.223083     & 22810$\pm$240               & 7.052$\pm$0.145      	& -3.070$\pm$0.056                & 9.3$\pm$7.2            & 23.1 &A\\
J0511+3917		& 77.896933 & 39.285150       & 39710$\pm$700               & 6.192$\pm$0.078      	& -0.151$\pm$0.091              & 4.8$\pm$13.6        & 11.3 &C\\
J0511+1408 		& 77.932619 & 14.136055     & 51170$\pm$1490              & 8.147$\pm$0.188       	& -2.822$\pm$0.155               & 8.8$\pm$23.5        & 17.0 &A\\
J0514+1137		& 78.620174 & 11.621418      & 33360$\pm$1370             & 5.661$\pm$0.163       	& -0.755$\pm$0.114               & 2.6$\pm$18.6        & 11.7 &C\\
J0517+2509 		& 79.322516 & 25.159017      & 68000$\pm$5000             & 8.500$\pm$0.500       	& -1.050$\pm$0.400              & 0.7$\pm$16.6          & 11.9 &D\\
J0524+1832 		& 81.217895 & 18.542563      & 50000$\pm$9500              & 6.481$\pm$0.544       	&$<$ -2.090          	             & 65.2$\pm$30.0      & 10.7 &D\\
J0526+0456 		& 81.590253 & 4.935078        & 71240$\pm$8290              & 6.887$\pm$0.335       	&$<$ -3.178          	             & 104.5$\pm$30.0    & 15.3 &C\\
J0529+0407 		& 82.391388 & 4.122033        & 34450$\pm$560                & 5.828$\pm$0.093       	& -0.958$\pm$0.058              & -31.8$\pm$19.3     & 19.4 &B\\
J0529+0407		& 82.391388 & 4.122033        & 36010$\pm$930                & 5.944$\pm$0.130         	& -0.922$\pm$0.106              & -21.3$\pm$18.7        & 12.6 &C\\
J0533+1606		& 83.338169 & 16.113207      & 34000$\pm$360                & 5.772$\pm$0.040         	& -1.616$\pm$0.031              & -182.9$\pm$16.9    & 58.5 &A\\
J0533+2126 		& 83.348132 & 21.438994      & 49590$\pm$3790              & 5.405$\pm$0.152       	& -3.409$\pm$0.161              & -6.9$\pm$7.8         & 27.5 &B\\
J0533+2126		& 83.348132 & 21.438994      & 42800$\pm$8340              & 5.538$\pm$0.720         	& -3.053$\pm$1.767              & 330.5$\pm$7.3      & 11.0 &D\\
J0539+6630 		& 84.996383 & 66.512067      & 42330$\pm$720                & 6.245$\pm$0.066       	& -2.380$\pm$0.126                & -52.5$\pm$21.6     & 16.9 &A\\
J0543+2504		& 85.899609 & 25.069656      & 69980$\pm$4150              & 7.303$\pm$0.333       	& -3.245$\pm$0.486              & 12.0$\pm$24.3        & 22.0 &C\\
J0543+2504		& 85.899609 & 25.069656      & 54310$\pm$3840              & 7.933$\pm$0.236       	&$<$ -3.139           	             & -22.3$\pm$15.0       & 37.7 &B\\
J0547+4008		& 86.775955 & 40.139606      & 32870$\pm$170                & 5.834$\pm$0.039           & -1.265$\pm$0.050                 & 1.3$\pm$4.9              & 37.7 &A\\
J0547+1709 		& 86.798328 & 17.155430        & 35180$\pm$760               & 5.830$\pm$0.129          & 1.690$\pm$0.472                  & 30.1$\pm$3.8            & 36.0 &C{\sc ii/iii},B\\
J0547+1709		& 86.798328 & 17.155430        & 35070$\pm$8350             & 5.233$\pm$0.871          & 1.442$\pm$0.323               & 22.6$\pm$13.5        & 11.4 &-\\
J0547$-$0024		& 86.817904 & -0.408748       & 14230$\pm$300                 & 6.566$\pm$0.105          & -0.802$\pm$0.254              & 18.9$\pm$21.9       & 17.3 &A\\
J0549+5222 		& 87.417267 & 52.374250        & 45730$\pm$820                 & 7.665$\pm$0.085          &$<$ -3.723           		     & -14.1$\pm$20.4      & 31.5 &A\\
J0556+0306		& 89.157363 & 3.101865        & 21140$\pm$1510               & 5.388$\pm$0.316           &$<$ -3.648                	     & 16.8$\pm$18.4       & 11.4 &A\\
J0609+0922		& 92.274279 & 9.371665        & 64890$\pm$5350               & 7.316$\pm$0.295        & -4.728$\pm$0.958               & 262.4$\pm$20.8      & 13.8 &C\\
J0609+4418		& 92.486533 & 44.308154      & 84390$\pm$4590               & 6.817$\pm$0.137        & -0.606$\pm$0.241                & 22.3$\pm$14.2        & 14.8 &B\\
J0616+0220		& 94.020607 & 2.337048        & 88930$\pm$6270               & 6.892$\pm$0.217        &$<$ -5.536           		       & 105.8$\pm$17.1       & 15.8 &B\\
J0618+2456 		& 94.659446 & 24.939967       & 29580$\pm$740                 & 5.454$\pm$0.115        & -2.395$\pm$0.100                     & -4.7$\pm$12.4         & 14.4 &A\\
J0618+2456		& 94.659446 & 24.939967      & 27140$\pm$1020               & 5.054$\pm$0.116        & -2.496$\pm$0.109                 & 15.5$\pm$8.2          & 16.2 &B\\
J0628+1120		& 97.126828 & 11.338466      & 42650$\pm$870                & 6.066$\pm$0.098        & 1.144$\pm$0.005                   & 43.8$\pm$16.2        & 10.3 &B\\
J0633+3720		& 98.425126 & 37.344831      & 32400$\pm$780                & 5.544$\pm$0.187        & -1.389$\pm$0.025                  & 29.0$\pm$10.2          & 18.7 &B\\
J0634+3803		& 98.707970 & 38.064487        & 21200$\pm$350                & 7.407$\pm$0.250          & -2.578$\pm$0.130                    & -60.0$\pm$7.5             & 14.1 &B\\
J0637+4222		& 99.308942 & 42.381702      & 34910$\pm$310                & 5.752$\pm$0.060          & -1.312$\pm$0.100                      & 22.3$\pm$8.2         & 27.1 & A\\
J0645+1140		& 101.383450 & 11.667548       & 25660$\pm$590              & 5.348$\pm$0.083        & -2.293$\pm$0.118                  & -84.3$\pm$9.0          & 29.5 &neb,A\\
J0645+1140 		& 101.383450 & 11.667548        & 26680$\pm$3110            & 5.385$\pm$0.367        & -2.346$\pm$0.263                  & -12.1$\pm$15.7       & 11.7 &B\\
J0647+1350 		& 101.895250 & 13.836488        & 31980$\pm$350              & 5.761$\pm$0.050          & -2.265$\pm$0.246                  & 36.9$\pm$5.4             & 37.1 &A\\
J0649+3457 		& 102.271810 & 34.960026        & 58710$\pm$4790            & 6.869$\pm$0.248        & -2.076$\pm$0.155                  & 32.8$\pm$24.6         & 13.5 &C\\
J0701+0941 		& 105.262300 & 9.693231            & 23120$\pm$340            & 5.422$\pm$0.050          & -2.598$\pm$0.697                 & 77.4$\pm$6.4           & 37.4 &A\\
J0707+1400		& 106.827510 & 14.016182       & 38800$\pm$220                & 5.627$\pm$0.034         & -1.940$\pm$0.027                   & 2.9$\pm$3.0              & 59.8 &A\\
J0707+1400		& 106.827510 & 14.016182       & 37940$\pm$750                & 5.806$\pm$0.147         & -1.646$\pm$0.057                 & -21.3$\pm$18.7       & 15.2 &A\\
J0710+0514 		& 107.713410 & 5.245842          & 34230$\pm$700               & 5.563$\pm$0.122          & -1.917$\pm$0.075                 & 76.7$\pm$19.8        & 12.2 &B\\
J0718+1026		& 109.734390 & 10.443990         & 31970$\pm$300               & 6.038$\pm$0.055          & -0.095$\pm$0.082                 & 11.7$\pm$3.8           & 64.6 &A\\
J0723+1140		& 110.862810 & 11.679968       & 27560$\pm$890               & 5.440$\pm$0.157             & -3.218$\pm$0.207                 & 111.9$\pm$12.6       & 19.1 &B\\
J0732+2704		& 113.083920 & 27.069038       & 37190$\pm$750               & 6.303$\pm$0.071           & 0.406$\pm$0.446                  & 92.8$\pm$11.8         & 16.8 & C{\sc iii},A\\
J0732+2704		& 113.084080 & 27.069423       & 38860$\pm$700               & 6.401$\pm$0.069           & 0.523$\pm$0.669                  & 115.0$\pm$8.8         & 13.3 &B\\
J0732+2704		& 113.084080 & 27.069423       & 29640$\pm$2010              & 5.214$\pm$0.185          & 1.383$\pm$0.001                  & 101.8$\pm$4.0         & 22.6 &C\\
J0737+4737		& 114.262030 & 47.630261       & 27670$\pm$410                & 7.486$\pm$0.112          &$<$ -4.699               		& 101.1$\pm$4.8         & 15.4 &A\\
J0737+0259		& 114.322805 & 2.998566       & 14580$\pm$450       	 & 4.667$\pm$0.103         &$<$ -2.771              		& 112.8$\pm$9.0           & 20.1 &comp,B\\
J0745+1949		& 116.298190 & 19.824048       & 7570$\pm$100            &$>$ 5.000           		 &-0.660$\pm$0.102       & -64.9$\pm$12.6          & 14.3 &BOSZ,met,B\\
J0745+2547		& 116.354210 & 25.785172      & 49450$\pm$3570         & 6.243$\pm$0.537            & -1.990$\pm$0.134                            & 41.2$\pm$12.1           & 25.3 &B\\
J0745+1858 		& 116.435090 & 18.980439       & 15570$\pm$320         & 7.347$\pm$0.271            &$<$ -1.878             			 & 34.2$\pm$19.0           & 16.6 &A\\
J0755+4800 		& 118.831120 & 48.009446       & 19150$\pm$240         & 7.245$\pm$0.072            &$<$ -2.103             			 & -147.9$\pm$12.0       & 17.4 &A\\
J0802+1750		& 120.678962 & 17.833711     & 22740$\pm$140        & 7.350$\pm$0.092               &$<$ -2.581             			 & -5.8$\pm$4.7            & 35.4 &A\\
J0803+4235 		& 120.891180 & 42.584640        & 25150$\pm$1440       & 6.369$\pm$0.191              & -2.997$\pm$0.090                            & 185.8$\pm$16.8       & 10.4 &B\\
J0804+3944		& 121.021510 & 39.743348       & 16410$\pm$410        & 7.333$\pm$0.127            & -1.819$\pm$1.168                         & -59.0$\pm$21.0        & 12.7&A\\
J0807+1321		& 121.913700 & 13.353081         & 17720$\pm$90          & 7.268$\pm$0.054            & -3.197$\pm$0.353                         & 17.7$\pm$3.1           & 55.8 &A\\
J0833+3852		& 128.473710 & 38.871847       & 35370$\pm$580        & 7.749$\pm$0.106             & -1.352$\pm$0.112                         & -18.4$\pm$14.2        & 13.7&DB,B\\
J0833+3852		& 128.473830 & 38.871903       & 34990$\pm$410        & 7.439$\pm$0.096            & -1.480$\pm$0.103                           & 29.0$\pm$11.7         & 22.7&DB,A\\
J0836+2057		& 129.082029 & 20.963603      & 30600$\pm$410      & 5.654$\pm$0.061             & -2.429$\pm$0.074                         & 89.5$\pm$8.7             & 31.9&A\\
J0905+3943		& 136.356200 & 39.727530     & 15560$\pm$620      & 7.210$\pm$0.261               &$<$ -1.920             			& 5.8$\pm$9.9            & 17.3 &B\\
J0919+2733		& 139.856010 & 27.550207      & 16350$\pm$360       & 7.347$\pm$0.148             &$<$ -1.889              			& 21.6$\pm$30.0         & 12.5&A\\
J0923+3028		 & 140.940058 & 30.467894     & 18270$\pm$90       & 6.812$\pm$0.026              &$<$ -2.738               			& -74.5$\pm$3.6            & 56.1&A\\
J0928+3016		 & 142.045201 & 30.267104      & 18010$\pm$5280   & 2.058$\pm$1.716             & -2.853$\pm$0.560               	  & 30.0$\pm$13.9         & 15.8&sdA?,C\\
J0935+4411		 & 143.778860 & 44.185285       & 20910$\pm$420      & 6.691$\pm$0.233             & -3.753$\pm$0.957                         & 30.0$\pm$20.2         & 12.2&C\\
J0935+4411		 & 143.778860 & 44.185285       & 21160$\pm$430      & 6.938$\pm$0.101              & -3.262$\pm$0.283                         & 22.0$\pm$5.4            & 15.5&A\\
J0937+3334		& 144.285863 & 33.567978     & 22680$\pm$310       & 7.340$\pm$0.097               &$<$ -2.767            			   & 66.6$\pm$20.5          & 14.8 &A\\
J0937+3334		 & 144.285863 & 33.567978      & 23910$\pm$200      & 7.197$\pm$0.051             & -3.216$\pm$0.109                        & -73.8$\pm$8.1         & 34.2&A\\
J0937+3334		 & 144.285682 & 33.567900         & 23520$\pm$100      & 7.249$\pm$0.071         & -3.095$\pm$0.075                        & 98.4$\pm$6.8          & 33.2 &A\\
J0938+2255		 & 144.599357 & 22.930363     & 17790$\pm$680      & 6.987$\pm$0.149             &$<$ -3.393            			  & 51.0$\pm$12.7        & 16.0 &A\\
J0947+2716		 & 146.872517 & 27.274167      & 26990$\pm$840      & 5.390$\pm$0.143              & -2.676$\pm$0.072                         & -71.4$\pm$9.3         & 14.8 &A\\
J0947+2716		 & 146.872334 & 27.274173     & 28340$\pm$460      & 5.531$\pm$0.091             & -2.913$\pm$0.030                          & 9.4$\pm$6.7             & 28.4 &A \\
J0950$-$0104		 & 147.582180 & -1.073202         & 63970$\pm$4590    & 7.248$\pm$0.379            & -1.787$\pm$0.751                        & 101.7$\pm$23.5       & 11.7 &C\\
J0950$-$0104		 & 147.582180 & -1.073202         & 69310$\pm$4280     & 6.934$\pm$0.070              & -2.105$\pm$0.121                        & 25.2$\pm$30.0           & 21.9 &B\\
J0950$-$0104		 & 147.582267 & -1.073167       & 61030$\pm$4200    & 7.538$\pm$0.198            & -1.837$\pm$0.168                        & 6.3$\pm$15.7             & 16.9 &B\\
J1005+2249		 & 151.496290 & 22.825639         & 20670$\pm$370       & 7.453$\pm$0.083           &$<$ -3.126            			  & 159.6$\pm$19.8      & 12.7 &A\\
J1005+2249		 & 151.496290 & 22.825639        & 19920$\pm$220       & 7.182$\pm$0.075           & -3.166$\pm$0.643                         & 88.2$\pm$12.9         & 18.9 &A\\
J1005+2249		 & 151.496290 & 22.825639        & 19610$\pm$280      & 7.078$\pm$0.067            & -3.132$\pm$0.362                         & 143.7$\pm$18.4       & 15.0 &A\\
J1009+0009		 & 152.480716 & 0.162220          & 17920$\pm$500       & 7.571$\pm$0.083            &$<$ -0.687            			   & 65.8$\pm$22.5         & 12.8 &A\\
J1013+2606		 & 153.425510 & 26.105560          & 50980$\pm$4110     & 5.722$\pm$0.141            & -1.890$\pm$0.106                          & 0.1$\pm$25.8           & 14.8 &B\\
J1013+2606		 & 153.425510 & 26.105560           & 50760$\pm$1750    & 5.486$\pm$0.113             & -1.706$\pm$0.124                        & -11.6$\pm$4.5            & 19.3 &A\\
J1013+2606		& 153.425510 & 26.105560           & 52690$\pm$4090     & 5.562$\pm$0.130              & -2.089$\pm$0.207                        & -34.5$\pm$5.8          & 37.1&B\\
J1015+0425		& 153.883740 & 4.418769          & 31550$\pm$530       & 7.569$\pm$0.116            &$<$ -3.417            			   & 75.1$\pm$30.0         & 10.3 &B\\
J1047+3453		 & 161.788200 & 34.896253          & 23560$\pm$1390     & 6.595$\pm$0.144            &$<$ -2.906              			   & -78.4$\pm$9.0           & 18.4&A\\
J1053+3156		 & 163.315170 & 31.944636        & 20080$\pm$300       & 6.941$\pm$0.113             &$<$ -3.006           			   & -6.3$\pm$4.9           & 20.4 &A\\
J1104+3610		 & 166.135542 & 36.180406      & 15620$\pm$70        & 7.450$\pm$0.042                &$<$ -1.915                         		   & -14.2$\pm$4.2           & 44.8&b,A\\
J1104+0918		 & 166.153104 & 9.306503        & 16620$\pm$220        & 7.474$\pm$0.082            & -2.242$\pm$0.518                         & -46.3$\pm$9.7          & 40.3&A\\
J1104+0918		 & 166.153104 & 9.306503        & 16250$\pm$220         & 7.386$\pm$0.031           & -2.176$\pm$0.353                         & -26.4$\pm$7.6         & 36.8&A\\
J1120+1559		 & 170.061762 & 15.998883      & 35900$\pm$380         & 7.081$\pm$0.086           &$<$ -2.363             			     & -22.3$\pm$15.1       & 20.4&A\\
J1129+4715		& 172.308833 & 47.250481      & 11670$\pm$50          & 5.307$\pm$0.056             &$<$ -4.952                           	     & -38.9$\pm$3.0            & 66.9&PET,met,A\\
J1129+4715		& 172.308833 & 47.250481       & 11150$\pm$90         & 5.055$\pm$0.017             &$<$ -4.634                           	     & 94.9$\pm$5.4           & 33.0&PET,A\\
J1133$-$0234		 & 173.334410 & -2.577587         & 25100$\pm$1770       & 7.133$\pm$0.247           &$<$ -2.734             			     & 69.0$\pm$20.8         & 11.6&B\\
J1144+3641		& 176.192358 & 36.697772       & 17120$\pm$380         & 7.775$\pm$0.112           &$<$ -1.848           			     & 6.5$\pm$22.0           & 10.9&B\\
J1150+0250		 & 177.596175 & 2.847963        & 16270$\pm$270         & 7.501$\pm$0.079           &$<$ -1.319                			     & 90.8$\pm$10.9         & 27.3&A\\
J1150+0250		 & 177.596175 & 2.847963        & 18740$\pm$470        & 7.537$\pm$0.168            &$<$ -2.158           			     & 50.3$\pm$14.5         & 11.6&B\\
J1235+1543		 & 188.957870 & 15.722032        & 19920$\pm$160        & 7.150$\pm$0.056              &$<$ -2.961           			     & -7.2$\pm$10.2           & 28.8 &A\\
J1242+4340		 & 190.507240 & 43.673145         & 37310$\pm$840        & 5.777$\pm$0.136            & -0.318$\pm$0.201                        & 11.1$\pm$8.2           & 13.3&A\\
J1242+4340		 & 190.507226 & 43.673147      & 35940$\pm$360        & 5.837$\pm$0.040              & -0.336$\pm$0.087                         & -8.1$\pm$3.8           & 37.1&C{\sc ii/iii},A\\
J1249+2626		& 192.431542 & 26.434608       & 13290$\pm$100       & 6.031$\pm$0.010             & -4.902 $\pm$0.268          		    & 42.6$\pm$7.7        & 16.2&PET,A\\
J1304+3129		& 196.164890 & 31.484677        & 36110$\pm$870        & 6.131$\pm$0.178             & -0.670$\pm$0.102                         & -24.3$\pm$26.7       & 12.0&B\\
J1305+3858		 & 196.339957 & 38.983327      & 40650$\pm$400        & 6.563$\pm$0.035            & -2.261$\pm$0.052                        & -15.0$\pm$9.7         & 44.8&A\\
J1329+1230		 & 202.355060 & 12.507073        & 14040$\pm$270        & 7.462$\pm$0.063            &$<$ -2.734          			   & 14.6$\pm$14.5            & 19.9&A\\
J1336+0646		& 204.140283 & 6.773950          & 17160$\pm$110         & 7.396$\pm$0.014           &$<$ -2.811            			   & -73.0$\pm$3.2          & 65.2&A\\
J1336+0646		 & 204.140283 & 6.773950         & 16660$\pm$290          & 7.375$\pm$0.089           &$<$ -1.639           			   & -52.6$\pm$5.7             & 40.8&A\\
J1337+3952		 & 204.355224 & 39.877280       & 23820$\pm$2100       & 6.289$\pm$0.213              &$<$ -2.558           			   & 21.9$\pm$12.0          & 17.9&B\\
J1337+3952 		& 204.355140 & 39.877390         & 22480$\pm$720           & 6.038$\pm$0.123             &$<$ -4.053           			   & 18.6$\pm$7.6             & 15.6&B\\
J1341+0454		 & 205.381170 & 4.912973        & 67660$\pm$3200         & 6.185$\pm$0.123             & -1.42$\pm$0.133                         & 67.4$\pm$11.8           & 22.9&B\\
J1343+0826		& 205.901850 & 8.444295            & 15900$\pm$1170     & 4.915$\pm$0.209            & -3.176$\pm$1.006                       & 351.6$\pm$6.7          & 15.5&B\\
J1353+1656		 & 208.426590 & 16.947604       & 20120$\pm$550         & 6.289$\pm$0.101             & -4.042$\pm$0.393                       & 19.8$\pm$8.4           & 16.7&A\\
J1353+1656		 & 208.426716 & 16.947526     & 24140$\pm$490         & 6.562$\pm$0.080              & -3.133$\pm$0.071                        & 100.9$\pm$3.6           & 15.6&A\\
J1355+1956		 & 208.801450 & 19.946000     & 16470$\pm$310         & 5.506$\pm$0.122           &$<$ -2.086                         		   & -90.0$\pm$3.0          & 56.1&comp,B\\
J1430+3001		 & 217.535290 & 30.019919       & 22640$\pm$780         & 6.677$\pm$0.147           & -3.421$\pm$0.513                        & 15.6$\pm$16.6         & 10.5&A\\
J1449+1717		 & 222.488127 & 17.291496     & 12590$\pm$100        & 5.927$\pm$0.010            & -4.840$\pm$0.205           		  & 132.4$\pm$8.8         & 17.0&PET,A\\
J1511+4048		 & 227.791153 & 40.800365      & 25600$\pm$990       & 6.270$\pm$0.108               & -3.422$\pm$0.269                        & 4.5$\pm$9.1            & 25.2&B\\
J1536+5013		 & 234.066050 & 50.230731       & 21610$\pm$1010      & 6.247$\pm$0.139            &$<$ -3.587             			   & 68.2$\pm$9.9        & 26.2&A\\
J1536+5013		 & 234.066363 & 50.230330       & 19520$\pm$450         & 6.029$\pm$0.127           & -3.642$\pm$0.337                          & 48.7$\pm$1.9        & 46.0&B\\
J1557+2823		 & 239.285340 & 28.393339        & 15820$\pm$330       & 7.198$\pm$0.086            &$<$ -1.506              			    & -31.4$\pm$28.6    & 10.5&A\\
J1557+2823		 & 239.285340 & 28.393339        & 15290$\pm$400       & 7.315$\pm$0.065            & -2.842$\pm$0.491                          & 106.4$\pm$13.5    & 21.4&A\\
J1626+1622		 & 246.603830 & 16.367105        & 14670$\pm$510       & 4.120$\pm$0.085             & -1.996$\pm$0.494                          & -131.7$\pm$13.5    & 11.8&B\\
J1805+1512		& 271.342820 & 15.200315        & 29160$\pm$740       & 5.350$\pm$0.071              & -2.905$\pm$0.100                               & -103.8$\pm$4.7     & 36.5&A\\
J1815+0642		& 273.887090 & 6.706773           & 25430$\pm$260       & 5.620$\pm$0.053             &$<$ -4.045              			    & 17.7$\pm$3.5        & 43.4&A\\
J1825+0633		& 276.447960 & 6.560357           & 29000$\pm$390       & 7.805$\pm$0.099              & -2.629$\pm$0.190                          & -5.0$\pm$13.5       & 14.5&B\\
J1827+1758		& 276.828600 & 17.970835          & 70740$\pm$2030     & 7.319$\pm$0.128              & -4.524$\pm$0.804                        & -43.4$\pm$14.5     & 46.6 &B\\
J1908+2617		 & 287.179310 & 26.289463        & 51530$\pm$720        & 7.981$\pm$0.105            & -3.811$\pm$0.217                          & -7.9$\pm$11.2       & 44.4&A\\
J1937+3606		& 294.486970 & 36.107181        & 25270$\pm$1180       & 5.703$\pm$0.156            & -2.490$\pm$0.085                            & -93.3$\pm$7.3      & 21.9&A\\
J2116+3746		& 319.052010 & 37.770494        & 10360$\pm$140         & 5.133$\pm$0.079            &$<$ -4.634                           		& 140.9$\pm$4.9     & 16.4 &PET,A\\
J2147+1651		& 326.869100 & 16.856723          & 68890$\pm$1140       & 8.497$\pm$0.095            & -0.226$\pm$0.163                          & -3.2$\pm$7.2          & 22.9 &A\\
J2149+3432		& 327.436630 & 34.545298        & 18190$\pm$550        & 7.331$\pm$0.150               &$<$ -3.422             				& 107.7$\pm$6.3      & 11.9 &B\\
J2151+1536		& 327.962990 & 15.614796        & 83070$\pm$4530      & 8.657$\pm$0.189             & -0.306$\pm$0.157                          & -51.5$\pm$13.3     & 26.7&pg1159,C{\sc iv},B\\
J2215+0518		 & 333.755760 & 5.309270           & 75480$\pm$3930       & 7.131$\pm$0.233            & -2.358$\pm$0.049                          & 109.2$\pm$15.7    & 16.2&B\\
J2215+0518		 & 333.755760 & 5.309264        & 65300$\pm$2770        & 7.666$\pm$0.188            &$<$ -3.251              				& 16.6$\pm$30.0     & 21.1 &A\\
J2240+1205		 & 340.088400 & 12.092550          & 30110$\pm$210          & 5.391$\pm$0.028             &$<$ -3.302             				& -167.3$\pm$5.0    & 39.7&A\\
J2251+3300		& 342.899090 & 33.013967       & 55900$\pm$3510       & 7.296$\pm$0.135              & -3.779$\pm$0.480                            & -20.1$\pm$29.1    & 30.3 &C\\
J2257+1216		 & 344.266849 & 12.279686    & 18530$\pm$290        & 6.932$\pm$0.084               & -2.990$\pm$0.553                           & 13.7$\pm$5.4         & 30.9&A\\
J2306+0224		 & 346.657764 & 2.408254       & 10810$\pm$70        & 5.126$\pm$0.062               &$<$ -4.912            				& 64.9$\pm$4.1          & 33.4&PET,A\\
J2323+1835		 & 350.904611 & 18.584587     & 25900$\pm$710       & 5.648$\pm$0.078               & -4.306$\pm$0.726                            & -0.4$\pm$9.6         & 22.4&A\\
J2323+4516		& 350.957890 & 45.281708       & 64880$\pm$3940       & 7.689$\pm$0.244             & -1.799$\pm$0.206                         & -6.8$\pm$30.0          & 16.5&A\\
J2331+4228		& 352.886960 & 42.474530         & 57420$\pm$630          & 7.343$\pm$0.047             & -3.230$\pm$0.067                           & -6.2$\pm$4.5            & 55.7&B\\
J2332+5056		& 353.233940 & 50.949957       & 14110$\pm$130          & 4.907$\pm$0.122             & -4.971$\pm$0.130                         & -156.5$\pm$9.0     & 26.9&PET,met,comp?,B\\
\enddata
\tablecomments{ $^{a}$ $y=\log{(n{\rm He}}/n{\rm H})$; ``$<$" denotes an upper limit of $y$ for the object; ``$>$" shows a lower limit of $\log{g}$ for the object.
}
\end{deluxetable*}

\subsection{Comparison to the ELM Survey}\label{sec23}
\begin{figure}
\center
\includegraphics[scale=0.7]{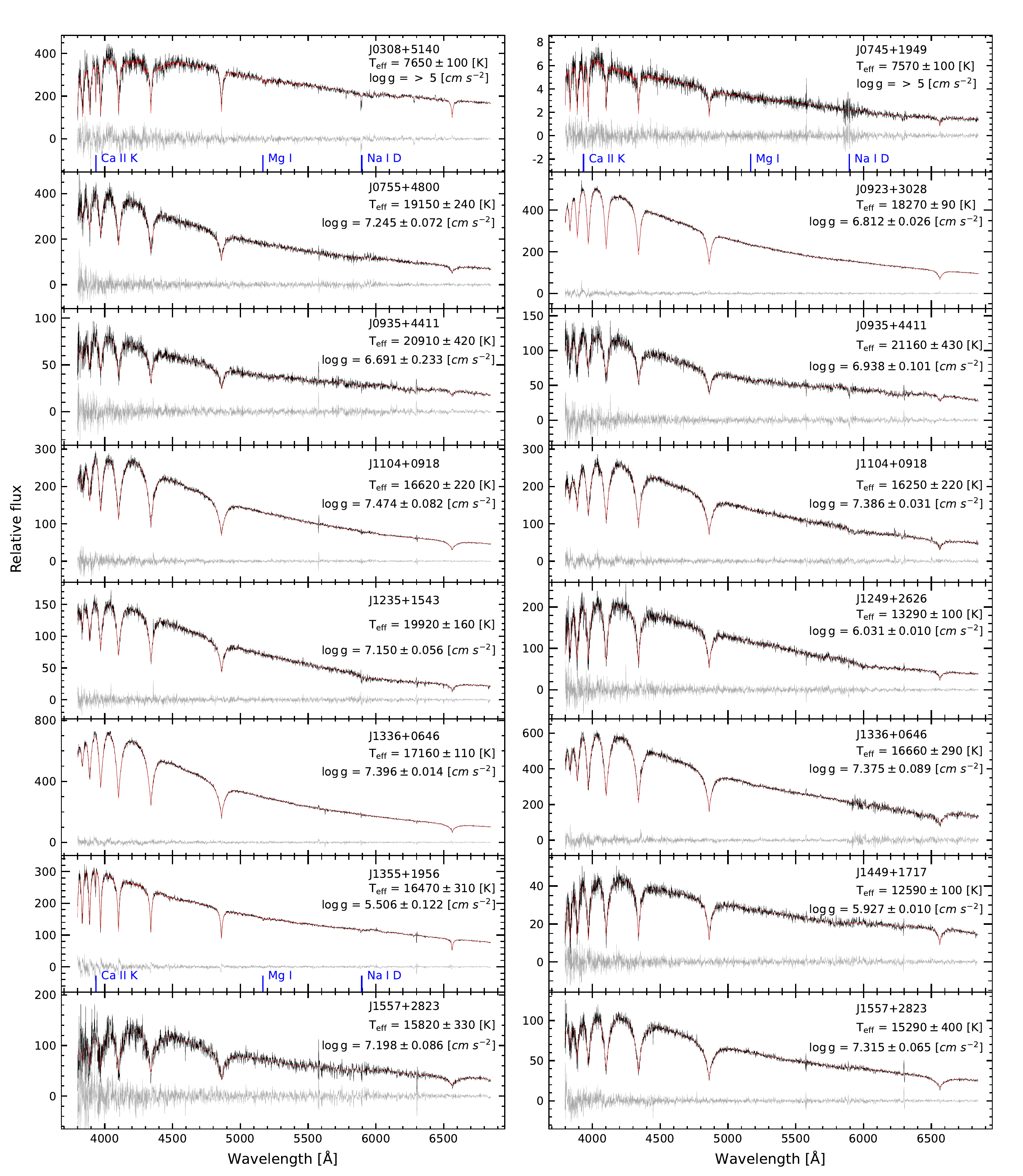}
\caption{Best model-atmosphere fits to the 16 LAMOST spectra of 12 stars in the ELM survey. 
The black and red lines are the observed and synthetic spectra, respectively. 
The residuals are shown as dark grey lines.
The vertical blue lines mark the locations of the Ca II K (3934.8 \AA), the Mg I (5167 \AA), and the Na I D (5895.6 \AA) absorption lines. 
The target name and the besting-fitting atmospheric parameters are given for each LAMOST spectrum.}
\label{fig2}
\end{figure}

\begin{figure}
\center
\includegraphics[scale=0.85]{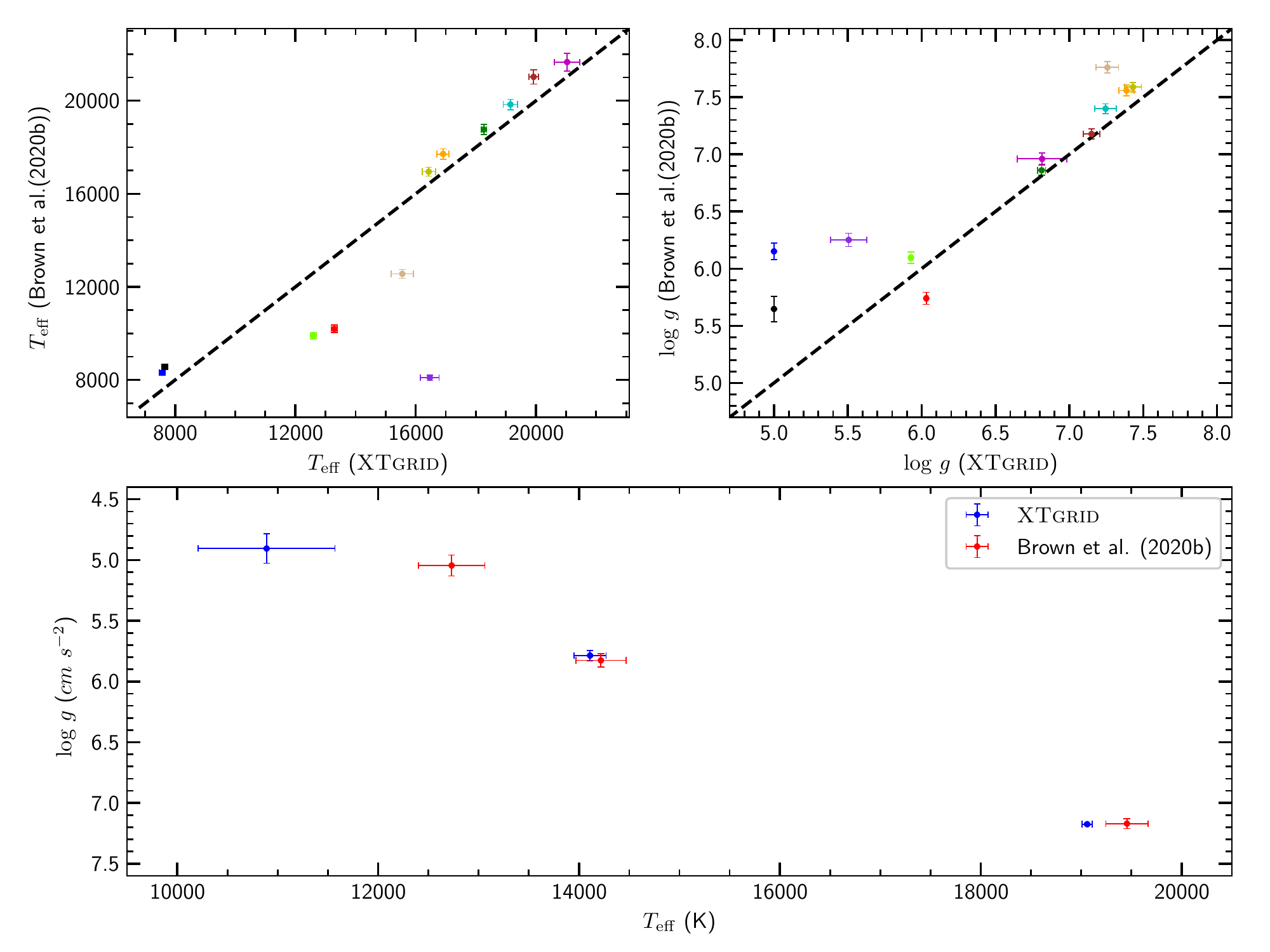}
\caption{Upper panel: comparisons of the atmospheric parameters measured by {\sc XTgrid} with LAMOST spectra 
and results of \cite{Brown2020} for the 12 known ELM WDs. We plot the same object with the same color in each of the upper panels.
The dashed lines are the 1:1 ratios.  
Lower panel: atmospheric parameters of three ELM WDs derived from two different fits to the same three MMT spectra (Warren R. Brown, private communication; \citealt{Brown2020}).
Our spectral fitting results and parameter values from \cite{Brown2020} are shown as blue and red points, respectively.}
\label{fig3}
\end{figure}

We found 12 stars in our sample that were also part of the ELM Survey. 
There are a total of 16 LAMOST spectra for the 12 stars, four of which have two LAMOST spectra. 
Figure \ref{fig2} shows our best model atmosphere fits to the 16 LAMOST spectra, 
where the black and red lines represent the observed and synthetic spectra, respectively. 
The residuals are displayed as dark grey lines. 
In Table \ref{tab2}, we have compared the atmospheric parameters of the stars from this work 
to the results published in the ELM Survey paper (see table 1 of \cite{Brown2020}). 
The effective temperature ($T_{\rm eff}$) and surface gravity ($\log{g}$) derived from our spectral fitting are listed in the fourth and fifth columns, respectively.
We have added notes in the sixth column to point out on which of the atmospheric models the listed parameters are based.
The last two columns give the corresponding results reported by \cite{Brown2020}. 
Parameter values from our spectral fitting and \cite{Brown2020} are also shown in Figure \ref{fig3}.
It is worth mentioning here that we have shown the average values of the parameters for the four stars in Table \ref{tab2} that have more than one spectrum.

We see that $T_{\rm eff}$ and $\log{g}$ from the two separate analyses match particularly well for hot stars with effective temperatures higher than 15\,000 K, 
while for cool stars ($T_{\rm eff}<9000$ K), our {\sc Tlusty} models yield notably higher $T_{\rm eff}$ than those from the ELM Survey.
That is probably because these stars are too cool for {\sc Tlusty} to produce self-consistent atmosphere models. 
The model convergence rate got progressively worse below 15\,000 K. 
Therefore, we modified the procedures for the cool stars, and applied the WD models developed by \cite{Tremblay11, Tremblay13, Tremblay15} for temperatures below 14\,000 K.
As seen in Table \ref{tab2}, our single star fits cannot fully reproduce the spectra for the 12 previously known objects, which may be due to binarity.
As depicted in Figure \ref{fig2}, the three cool stars ($T_{\rm eff}<9000$ K) from the ELM Survey show strong metal lines, suggesting they may be composite binary systems with a cool component. 
The two cool objects in Table\,\ref{tab2}, J0308+5140 and J0745+1949, show metal lines and our {\sc Tlusty} models were unable to reproduce their spectra.
We applied pre-calculated {\sc Atlas9} LTE models from the BOSZ \citep{bohlin17} spectral library. 
Unfortunately, the BOSZ library does not cover the parameter space over $\log{g}=5$. 
These stars are most consistent with subdwarf A-type stars (sdAs).
sdA stars have hydrogen-rich spectra and surface gravities between those of main-sequence stars and isolated WDs, 
but their effective temperatures are below the zero-age horizontal branch \citep{Kepler2016, bell18, Yu2019}.
J1355+1956 is a composite spectrum binary, which we decomposed into its components, and here we report the parameters only for the hot star.
However, in the ELM Survey it is listed as a pulsating single sdA star \citep{Bell2017, Brown2020}.
Further spectroscopic and photometric observations could be helpful to reveal the nature of J1355+1956.
Disregarding these binary candidates, Table\,\ref{tab2} reflects that the systematics in $T_{\rm eff}$ towards low-temperature stars is not extreme. 
It is also important to note that we have derived the parameters from spectra of different resolution, coverage, and quality.   

In order to test the validity of our spectral fitting method further, 
we reanalyzed three MMT spectra (Warren R. Brown, private communication; \citealt{Brown2020}) to see if we could recover the same parameters.
The results, derived from the three MMT spectra, are given in the last three rows of Table \ref{tab2} and are also shown in the lower panel of Figure\,\ref{fig3}. 
We note offsets between the two analyses, which can be due to the different atmosphere models used.
It should be noted here that the upper panel of Figure \ref{fig3} seems to suggest that 
the problem with the atmospheric fit is not just with lower effective temperatures, but also lower $\log{g}$ values.
However, from Table \ref{tab2} it happens that the WDs with lower $\log{g}$ values also have lower $T_{\rm eff}$ values, 
and the comparison of the fitting of the three MMT spectra shows that the problem is indeed most likely with the temperature.
 
\begin{deluxetable}{lrrrrrrrr}
\tablecolumns{9}
\tablecaption{Comparison of the atmospheric parameters from this work to the results published in ELM survey paper (see Table 1 of \cite{Brown2020}). 
The effective temperature ($T_{\rm eff}$) and surface gravity ($\log{g}$) obtained by {\sc XTgrid} are listed in the fourth and fifth columns, respectively.
To point out on which of the atmospheric models the listed parameters are based, we added notes in the sixth column.
The last two columns give the results from \cite{Brown2020}.\label{tab2}}
\tabletypesize{\scriptsize}
\tablewidth{0pc}
\tablehead{
\colhead{}  &\colhead{}  &\colhead{}  & \multicolumn{3}{c}{{\sc XTgrid}}  &\colhead{} & \multicolumn{2}{c}{\cite{Brown2020}} \\
\cline{4-6} \cline{8-9}
\colhead{Object} & \colhead{RA$_{J2000}$} & \colhead{DEC$_{J2000}$} & \colhead{$T_{\rm eff}$} & 
\colhead{$\log{g}$}  &\colhead{Notes}  &\colhead{}   & \colhead{$T_{\rm eff}$} & \colhead{$\log{g}$}	 \\
\colhead{} &\colhead{(degree)}      &\colhead{(degree)}  &\colhead{($\rm K$)}     
&\colhead{($\rm cm~s^{-2}$)}  &\colhead{}  &\colhead{}    &\colhead{($\rm K$)}   &\colhead{($\rm cm~s^{-2}$)}	
}
\startdata
J0308+5140  			& 47.075785  	& 51.669867  	& 7650$\pm$100     		&$>$ 5.000     			&BOSZ     	&& 8559$\pm$110      & 5.647$\pm$0.112\\
J0745+1949  			& 116.298190  	& 19.824048  	& 7570$\pm$100      		&$>$ 5.000     			&BOSZ		&& 8313$\pm$100      & 6.151$\pm$0.074	\\
J0755+4800  			& 118.831120  	& 48.009446  	& 19150$\pm$240    		& 7.245$\pm$0.072     	&{\sc Tlusty}	&& 19839$\pm$220     & 7.399$\pm$0.043	\\
J0923+3028   			& 140.940058 	& 30.467894  	& 18270$\pm$90     		& 6.812$\pm$0.026     	&{\sc Tlusty}	&& 18761$\pm$220     & 6.860$\pm$0.044	\\
J0935+4411   			& 143.778860  	& 44.185285  	& 20910$\pm$420    		& 6.691$\pm$0.233     	&{\sc Tlusty}	&& 21660$\pm$380     & 6.960$\pm$0.050	\\
J0935+4411   			& 143.778860  	& 44.185285  	& 21160$\pm$430    		& 6.938$\pm$0.101     	&{\sc Tlusty}	&& 21660$\pm$380     & 6.960$\pm$0.050	\\
J1104+0918   			& 166.153104 	& 9.306503   	& 16620$\pm$220    		& 7.474$\pm$0.082     	&{\sc Tlusty}	&& 16952$\pm$190     & 7.588$\pm$0.042	\\
J1104+0918   			& 166.153104 	& 9.306503   	& 16250$\pm$220    		& 7.386$\pm$0.031     	&{\sc Tlusty}	&& 16952$\pm$190     & 7.588$\pm$0.042	\\
J1235+1543   			& 188.957870  	& 15.722032  	& 19920$\pm$160    		& 7.150$\pm$0.056     	&{\sc Tlusty}	&& 21024$\pm$310     & 7.178$\pm$0.044	\\
J1249+2626     			& 192.431542 	& 26.434608  	& 13290$\pm$100    		& 6.031$\pm$0.010     	&PET		&& 10195$\pm$160     & 5.740$\pm$0.053\\
J1336+0646     			& 204.140283 	& 6.773950 	& 17160$\pm$110    		& 7.396$\pm$0.014     	&{\sc Tlusty}	&& 17710$\pm$220     & 7.558$\pm$0.046	\\
J1336+0646   			& 204.140283 	& 6.773950    	& 16660$\pm$290      	& 7.375$\pm$0.089     	&{\sc Tlusty}	&& 17710$\pm$220     & 7.558$\pm$0.046	\\
J1355+1956	   		& 208.801450  	& 19.946000 	& 16470$\pm$310       	& 5.506$\pm$0.122     	&{\sc Tlusty}	&& 8096$\pm$120	 & 6.251$\pm$0.058	\\
J1449+1717  			& 222.488127 	& 17.291496  	& 12590$\pm$100     	& 5.927$\pm$0.010     	&PET		&& 9913$\pm$140	 & 6.095$\pm$0.050	\\
J1557+2823   			& 239.285340  & 28.393339  	& 15820$\pm$330      	& 7.198$\pm$0.086     	&{\sc Tlusty}	&& 12560$\pm$190     & 7.760$\pm$0.050	\\
J1557+2823   			& 239.285340  & 28.393339  	& 15290$\pm$400       	& 7.315$\pm$0.065     	&{\sc Tlusty}	&& 12560$\pm$190     & 7.760$\pm$0.050\\
J0050$+$2147$^{a}$	& 12.695213  	& 21.790461  	& 14110$\pm$160	   	& 5.786$\pm$0.042	 	&{\sc Tlusty}	&& 14218$\pm$250	 & 5.826$\pm$0.053	\\
J0441$-$0547$^{a}$		& 70.385937  	&-05.793042  	& 10890$\pm$680    		& 4.905$\pm$0.122	 	&PET		&& 12732$\pm$330	 & 5.045$\pm$0.086	\\
J0923$-$1218$^{a}$		& 140.959662 	&-12.306667	& 19060$\pm$50	   	& 7.175$\pm$0.010	 	&{\sc Tlusty}	&& 19455$\pm$210	 & 7.170$\pm$0.041	\\
\enddata
\tablecomments{$^a$ MMT spectra provided by Dr. Warren R. Brown \citep{Brown2020}; ``$>$" denotes a lower limit of $\log{g}$ for the object.}
\end{deluxetable}

\section{Stellar Parameters and  Parallaxes}\label{section3}
\begin{figure}
\center
\includegraphics[scale=0.8]{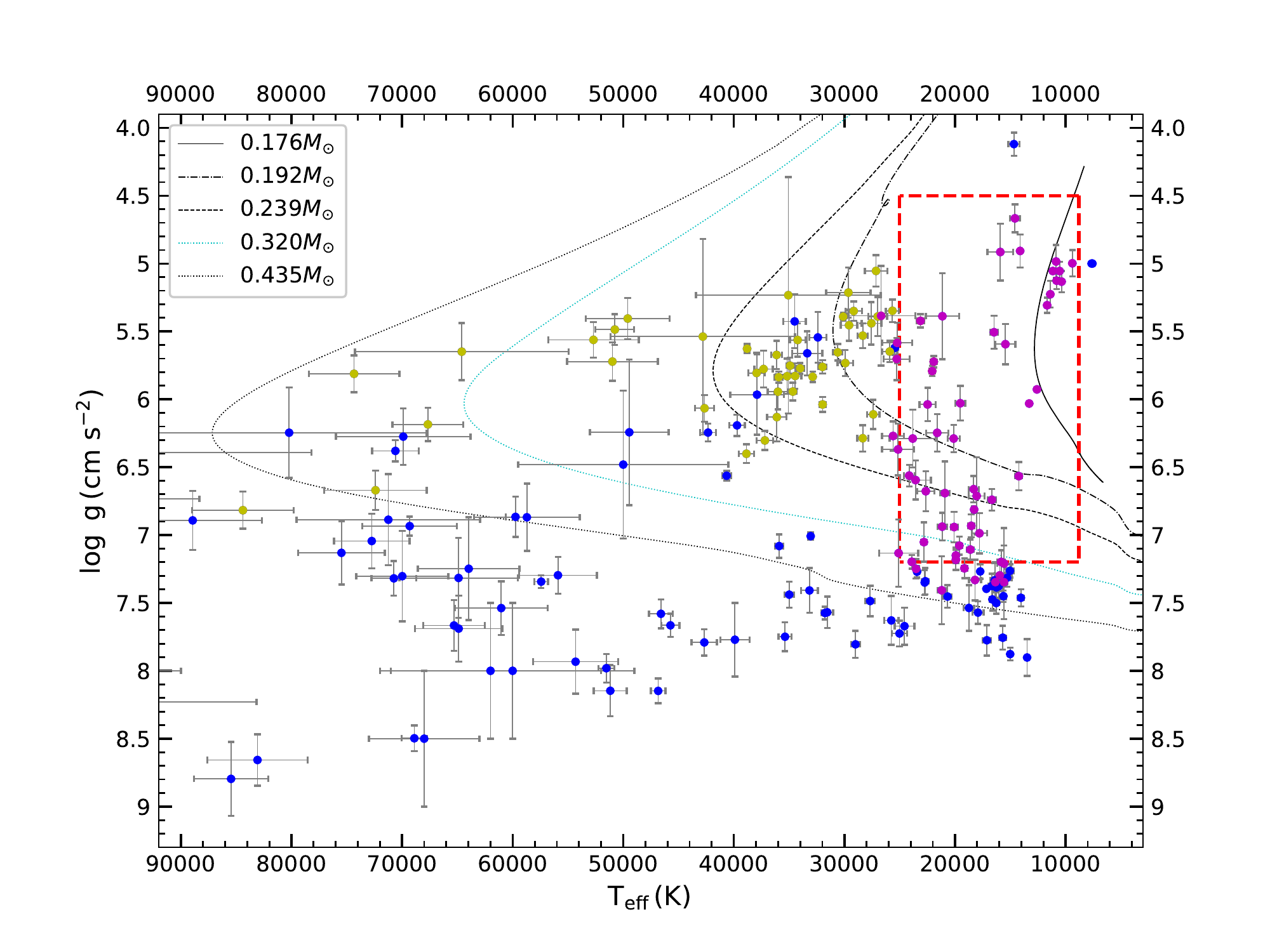}
\caption{$T_{\rm eff}$ vs. $\log{g}$ diagram of all the 188 LAMOST spectra. Evolutionary tracks from \cite{Althaus2013} for low mass WDs with 0.17$M_{\sun} < M < 0.44M_{\sun}$ are overplotted.
The magenta dots mark the selected ELM WD candidates that fall in the red box with effective temperatures of 8800 K $\leq$ $T_{\rm eff}$ $\leq$ 25000 K and surface gravities of 4.5 $\leq$ $\log{g}$ $\leq$ 7.2.
The yellow dots represent the hot subdwarf stars identified by \cite{Luo2021}.}
\label{fig4}
\end{figure}

Figure \ref{fig4} displays the distributions of the effective temperature, $T_{\rm eff}$, and surface gravity, $\log{g}$, for all the 188 LAMOST spectra. 
Evolutionary tracks from \cite{Althaus2013} for low mass WDs with 0.17$M_{\sun} < M < 0.44M_{\sun}$ are marked. 
The distributions of the points in Figure \ref{fig4} indicates a large variety of stars in our sample.
We cross-matched our results with the latest catalog of hot subdwarf stars \citep{Luo2021} 
and found 36 objects to be hot subdwarf stars, which are shown as yellow dots in Figure \ref{fig4}. 
Helium-burning sdB stars are mainly found in the range 25000 K$ < T_{\rm eff} < $ 40000 K and 5 $ < \log{g} < $ 6 \citep{Heber2009}. 
The \cite{Althaus2013} tracks suggest low surface gravities ($\log{g}$ $\lesssim$ 7.2) for ELM WDs. 
Accordingly, we picked out these objects in the region 8800 K$ \leq T_{\rm eff} \leq $ 25000 K and 4.5 $\leq$ $\log{g}$ $\leq$ 7.2 as ELM WD candidates.
A total of 58 LAMOST spectra, illustrated as magenta dots in Figure \ref{fig4}, were chosen for further analysis.
In order to create a clean sample of ELM WDs, \cite{Brown2020} applied stricter cuts with effective temperatures of 8800 K$ < T_{\rm eff} < $ 22000 K 
and surface gravities of 5.5 $\leq \log{g} \leq$ 7.1.
If these stricter criteria were applied, the number of candidates found here would be 28. 
However, as we sought to find as many ELM WDs candidates and their precursors as possible, we used a larger range of $T_{\rm eff}$ and $\log{g}$.   

\subsection{Stellar mass, radius, and luminosity} 
 Following the approach of \citet{Brown2020}, the mass of every ELM WD candidate was estimated 
 by interpolating its effective temperature and surface gravity with \citet{Althaus2013}'s evolutionary tracks for ELM WDs. 
 The radii of these objects were given directly by the stellar mass and surface gravity, 
 \begin{equation} 
 \log{g} = \log{g_{\odot}} + \log{ (M / R^2) }, 
 \end{equation} 
 where the solar surface gravity is $\log{g_{\odot}}$ = 4.4374 $\pm$ 0.0005 \citep{Gray1992}, where $g$ is in units of cm s$^{-2}$, 
 and $M$ and $R$ are in units of solar mass and radius, respectively. The luminosity of the star was then calculated with the relation $\frac{L}{L_{\sun}}$ = $(\frac{T}{T_{\sun}})^4$$(\frac{R}{R_{\sun}})^2$, 
 where the solar effective temperature is $T_{\sun}$ = 5777$\pm$10 K \citep{Gray1992}. The uncertainties of these parameters were derived via Monte Carlo error propagation. 
 As has been pointed out by \cite{Brown2020}, the WD masses obtained from \cite{Althaus2013} and \cite{Istrate2016} tracks differ by 0.000 $\pm$ 0.012$M_{\sun}$. 
 We thus added $\pm$0.01$M_{\sun}$ in quadrature to the mass errors, similar to \cite{Brown2020}. 
 
\subsection{Parallax} 
To cross-check the nature of the selected objects, we converted the luminosities into spectrophotometric parallaxes 
for comparison with the precise astrometric parallaxes available from $\it Gaia$ EDR3 \citep{Gaia2021}.
The absolute magnitude ($G_{abs}$) for each star was derived using
\begin{equation} 
G_{abs} = -2.5\log{L} + M_{bolo, \sun} - BC_{G}, 
 \end{equation}
 where $L$ is given in solar units and $M_{bolo, \sun}$ = 4.74. The bolometric correction ($BC_{G}$) for every candidate was obtained 
 by interpolating the $\it MIST$ bolometric correction tables\footnote{http://waps.cfa.harvard.edu/MIST/model\_grids.html\#bolometric} with solar metallicity \citep{Dotter2016, Choi2016}.
 The spectrophotometric parallaxes, $\pi_{spec}$, were then calculated with the following expression.
 \begin{equation} 
G_{abs} = G + 5 +5\log{\pi_{spec}} - A_{G}, 
 \end{equation}
 where $G$ is the $\it Gaia$ EDR3 $G$-band magnitude, the parallax, $\pi_{spec}$, is given in arcseconds, and $A_{G}$ represent interstellar extinction.
 We obtained $E(B-V)$ from Green's 3D dust map \citep{Green2019} and converted it to $A_{G}$ by 
 using the WC2019 extinction law \citep{Wangchen2019}, $E$($G_{BP}$ - $G_{RP}$) = $E(B-V)$ / 0.757 and $A_{G}$ = 1.890$E$($G_{BP}$ - $G_{RP}$).
 
 Based on the positions, we cross-matched the 58 selected LAMOST spectra against the $\it Gaia$ EDR3 catalog and found trigonometric parallaxes for all of them.
The zero point correction package for the $\it Gaia$ EDR3 parallaxes, developed by \cite{Lindegren2021}, was adopted to correct the parallaxes. 
The corrected $\it Gaia$ EDR3 parallaxes were found to have only a small systematic bias ($<$10 $\mu$as as \cite{Ren2021}).

\section{Results and Discussion}\label{section4} 
\begin{figure}
\center
\includegraphics[scale=0.75]{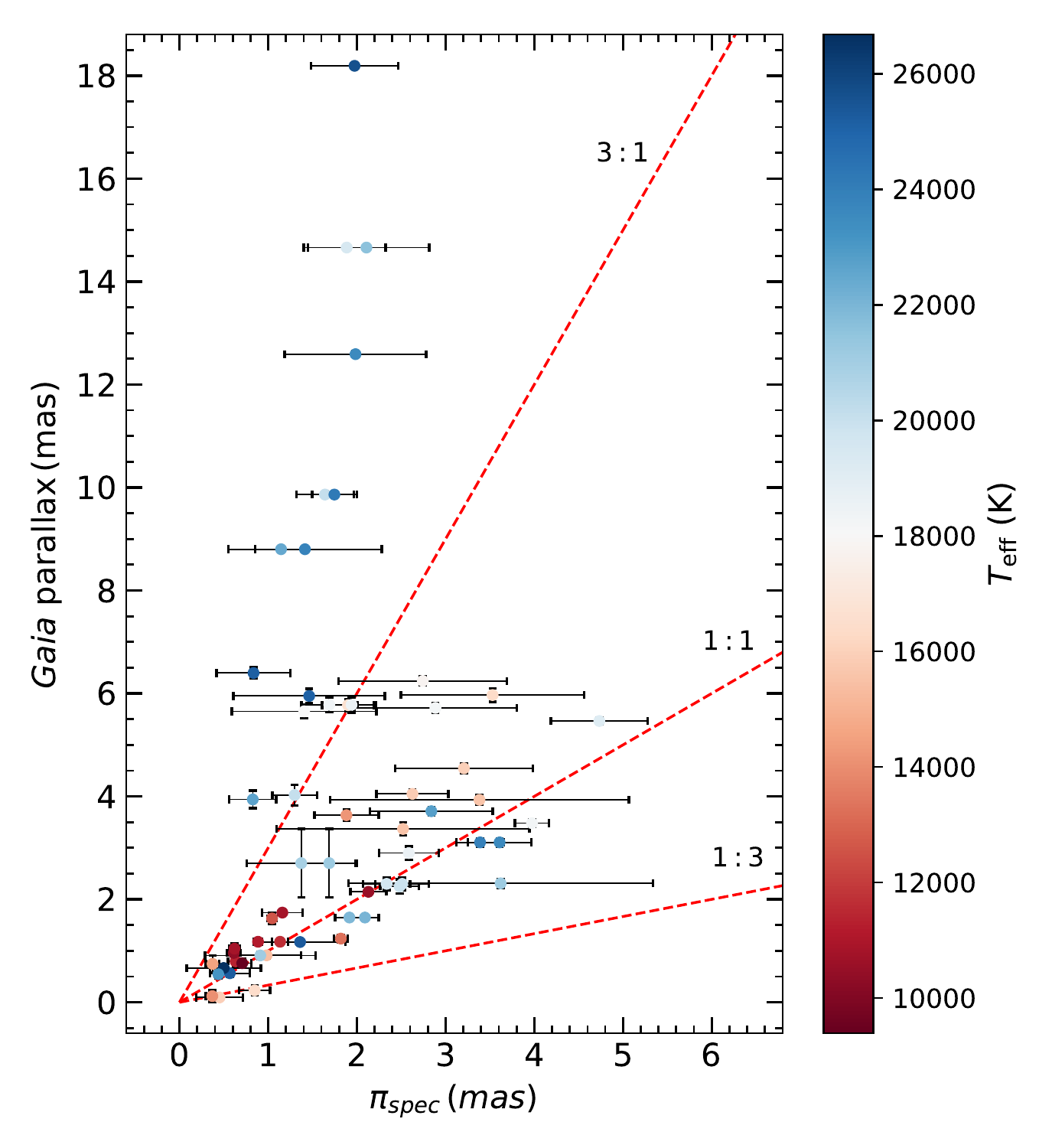}
\caption{Comparison between the $\it Gaia$ parallaxes and our spectrophotometric parallax estimates.
Red dashed lines represent the 3:1, 1:1, and 1:3 parallax ratio lines.
The marker colors indicate the effective temperatures of the objects.}
\label{fig5}
\end{figure}

Figure \ref{fig5} displays a comparison between the $\it Gaia$ parallaxes and our spectrophotometric parallax estimates.  
The red dashed lines mark the 3:1, 1:1, and 1:3 parallax ratio lines.
We note that most candidates have distance estimates that differ by less than three times. 
There are also some candidates clearly above the 3:1 ratio line.
However, none of the systems with $\pi$/$\sigma_{\pi}$ $\geq$ 5 are below the 1:3 ratio line.
\cite{Brown2020} showed a similar plot but with many more candidates from the ELM Survey (see their figure 3).
In \cite{Brown2020} most of the systems that did not follow the 1:1 ratio were present near the 1:30 ratio line where the $\it Gaia$ distance was much larger than the spectroscopic distance.
The authors argued that these objects near the 1:30 ratio line were attributed to sdA stars. 
However, here, most of the systems that do not follow the 1:1 ratio have smaller $\it Gaia$ distances compared to the spectroscopic ones, as shown in Figure \ref{fig5}.
More specifically, there are a total of 12 candidates above the 3:1 ratio line (i.e., J0043+2850, J0439+3954, J0458+1740, J0803+4235, J1047+3453, 
J1053+3156, J1133-0234, J1337+3952, J1353+1656, J1430+3001,  J1511+4048, and J1536+5013).
As can be seen from Table \ref{tab1}, the effective temperatures of the 12 objects are between 16660 and 25600 K, 
and their surface gravities range from $\log{g}$ $\sim$ 6.03 to 7.13. 
In order to find more information, we performed a detailed literature survey for each system above the 3:1 ratio line.
Nine of the 12 objects were found to have atmospheric parameters with 8510 K $\leq$ $T_{\rm eff}$ $\leq$ 9620 K and 7.75 $\leq$ $\log{g}$ $\leq$ 8.26 \citep{Tremblay11, Anguiano2017}.
As mentioned in Section \ref{sec23}, {\sc Tlusty} model atmospheres may not be well suited for cool stars with $T_{\rm eff}<15000$ K.
The {\sc XTgrid} fits based on {\sc Tlusty} model atmospheres might have overestimated the effective temperatures of the objects above the 3:1 line 
and yet underestimated their surface gravities.  
The estimated $L$ for a star above the 3:1 line, obtained by using Equation (1) and the relation of $\frac{L}{L_{\sun}}$ = $(\frac{T}{T_{\sun}})^4$$(\frac{R}{R_{\sun}})^2$, is likely larger than the true $L$. 
According to Equations (2) and (3), an overestimated luminosity may result in a smaller spectroscopic parallax compared to the $\it Gaia$ parallax.
This might suggest that these systems above the 3:1 line could be poorly fitted objects with lower effective temperatures, 
which are too cool for {\sc Tlusty} to produce self-consistent atmosphere models.      

We identified high-probability ELM WDs as objects with masses $M$ $\leq$ 0.3$M_{\sun}$ and whose parallax estimates agreed to within a factor of 3.
Excluding the previously known ELM WDs published by \cite{Brown2020}, there are a total of 21 new high-probability ELM WDs.
We have presented a summary of their physical parameters in Table \ref{tab3}, including target name,  $\it Gaia$ parallax, and our spectrophotometric parallax estimate.
Two of them, J0338+4134 and J1129+4715, have two LAMOST spectra and thus we have listed the average value of the parameters for the two stars. 
By fitting the two LAMOST spectra of J0338+4134, we obtained two RV measurements, RV$_{1}$ = -104.4$\pm$3.1 km/s and RV$_{2}$ = 79.4$\pm$3.3 km/s.
The fitting results for J1129+4715 were RV$_{1}$ = -38.9$\pm$3.0 km/s and RV$_{2}$ = 94.9$\pm$5.4 km/s.
We also note that the ratio of the $\it Gaia$ parallax to the predicted parallax for each of the two stars is close to 1:1, 
implying that they are strong ELM WD candidates.
By showing significant RV variability, J0338+4134 and J1129+4715 are very likely to be binary systems containing at least one ELM WD.
In addition, we measured a high projected rotation for J0338+4134, which may be the result of rotation, line blending, or orbital smearing.
Further high-resolution follow-up spectroscopic observations are required to confirm this measurement.

\begin{figure}
\center
\includegraphics[scale=0.7]{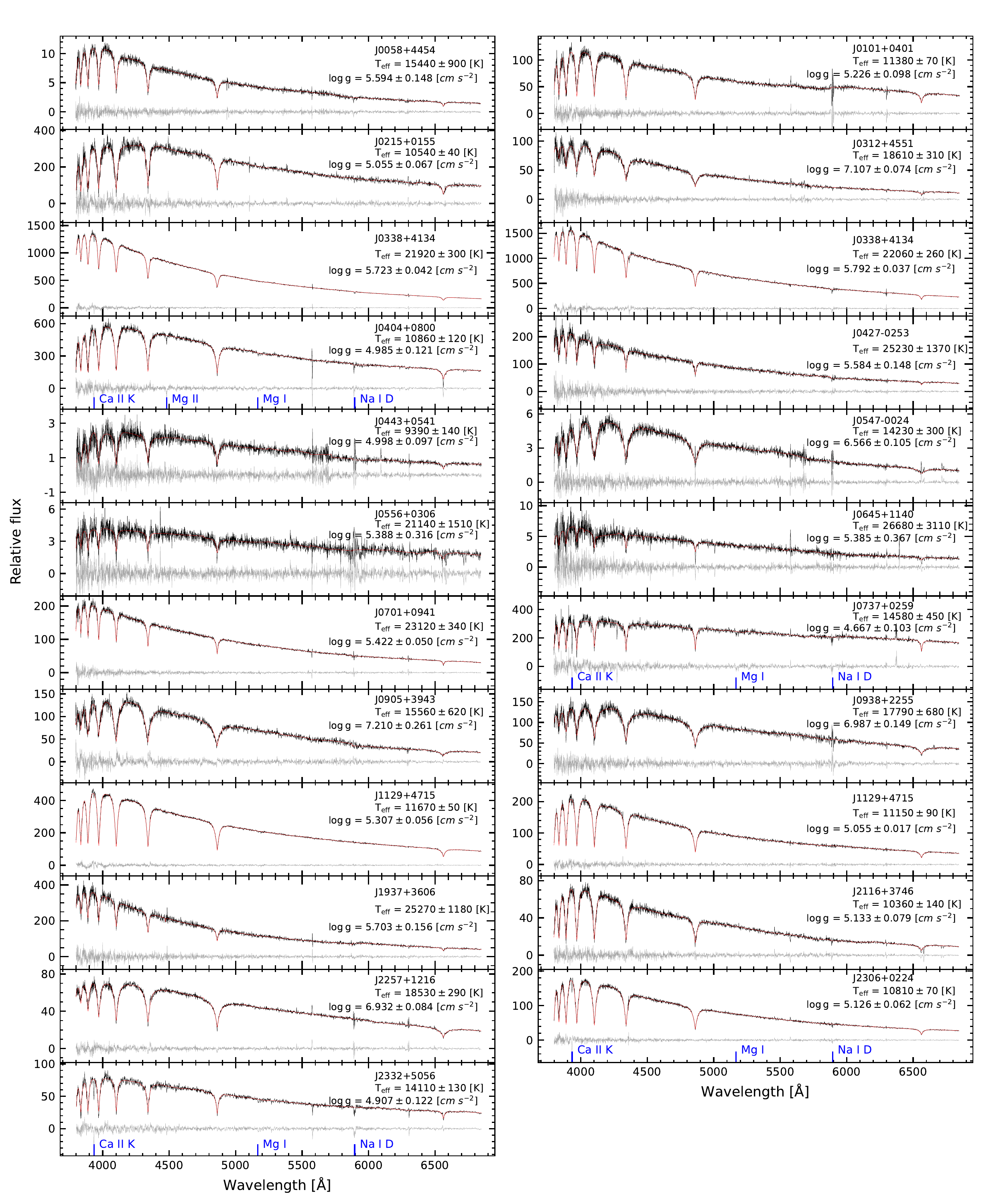}
\caption{Best model-atmosphere fits to the 23 LAMOST spectra for the 21 new high-probability ELM WDs. 
The black and red lines are the observed and synthetic spectra, respectively. The residuals are shown as dark-grey lines.
The vertical blue lines mark the locations of the Ca II K (3934.8 \AA), the Mg II (4481 \AA), the Mg I (5167 \AA), and the Na D (5895.6 \AA) absorption lines. 
The target name and the besting-fitting atmospheric parameters are given for each spectrum.}
\label{fig6}
\end{figure}

Figure \ref{fig6} shows our best model atmosphere fits to the 21 new high-probability ELM WDs. 
The observed and synthetic spectra are depicted as black and red lines, respectively. 
We can see that several relatively cool stars ($T_{\rm eff}$ $<$ 15000 K), such as J0404+0800, J0737+0259, and J2332+5056, present strong calcium, sodium, and even magnesium lines. 
The origin of these absorption lines might be interstellar, photospheric (e.g. \cite{Istrate2016}), or from a cool companion star. 
The RVs of the metal lines are consistent with those of the respective WD, within the measurement uncertainties.
Three candidates, J0427$-$0253, J0645+1140, and J1937+3606, have a higher effective temperature ($T_{\rm eff}$ $\simeq$ 25000 K) 
than the clean ELM WD sample defined by \cite{Brown2020}. 
However, the predicted parallax for each of these three stars agrees well with the $\it Gaia$ parallax.
We thus list them as high-probability ELM WDs.
Better spectroscopic data are needed to probe the nature of these stars.

In Table \ref{tab1}, J0833+3852 is listed as a DB star.
However, \cite{Tremblay11} suggested that J0833+3852 might be a DA+DB binary.
Future high-resolution spectra are required to reveal the nature of the system.    

\begin{deluxetable}{lrrrrrrrr}
\tablecolumns{9}
\tablecaption{The physical parameters of the 21 high-probability ELM WDs. \label{tab3}}
\tabletypesize{\scriptsize}
\tablewidth{0pc}
\tablehead{\colhead{Object} & \colhead{RA$_{J2000}$}  & \colhead{DEC$_{J2000}$} & \colhead{$T_{\rm eff}$} & \colhead{$\log{g}$}  & \colhead{Mass} & \colhead{Radius}  
& \colhead{Parallax$_{Gaia}$} & \colhead{Parallax$_{Spec}$}\\
\colhead{}  &\colhead{(degree)} &\colhead{(degree)} &\colhead{($\rm K$)}  &\colhead{($\rm cm~s^{-2}$)}  &\colhead{($M_{\sun}$)}  
&\colhead{($R_{\sun}$)}  &\colhead{(mas)}  &\colhead{(mas)}
}
\startdata
J0058+4454 	 & 14.654559   & 44.908567 	& 15440$\pm$900 & 5.594$\pm$0.148      &0.192$\pm$0.015 & 0.116$\pm$0.021 & 0.9305$\pm$0.0663 & 0.9822$\pm$0.3895\\
J0101+0401 	 & 15.369537   & 4.033043   	& 11380$\pm$70   & 5.226$\pm$0.098      &0.170$\pm$0.012 & 0.166$\pm$0.020 & 0.8266$\pm$0.0981 & 0.6327$\pm$0.0875\\
J0215+0155 	 & 33.776147   & 1.917815   	& 10540$\pm$40   & 5.055$\pm$0.067      &0.167$\pm$0.012 & 0.201$\pm$0.017 & 2.1889$\pm$0.0283 & 2.1298$\pm$0.2025\\
J0312+4551 	 & 48.085425   & 45.856474    	& 18610$\pm$310 & 7.107$\pm$0.074      &0.326$\pm$0.027 & 0.026$\pm$0.003 & 2.9098$\pm$0.1285 & 2.5853$\pm$0.3369\\
J0338+4134 	 & 54.696112   & 41.573393    	& 21990$\pm$280 & 5.758$\pm$0.040       &0.209$\pm$0.011 & 0.100$\pm$0.005 & 1.6778$\pm$0.0318 & 2.0034$\pm$0.1595\\
J0404+0800 	 & 61.076658   & 8.000846      	& 10860$\pm$120 & 4.985$\pm$0.121 	 &0.172$\pm$0.014 & 0.221$\pm$0.032 & 1.7848$\pm$0.0435 & 1.1602$\pm$0.2292\\
J0427$-$0253 	 & 66.750345   & -2.883789    	& 25230$\pm$1370 & 5.584$\pm$0.148     &0.215$\pm$0.018 & 0.124$\pm$0.022 & 0.5699$\pm$0.0714 & 0.5676$\pm$0.2248\\
J0443+0541	 & 70.761091   & 5.688080     	&9390$\pm$140   &4.998$\pm$0.097 	     &0.159$\pm$0.013 & 0.209$\pm$0.025 & 0.7917$\pm$0.0849 &0.7078$\pm$0.1036\\
J0547$-$0024	 & 86.817904   & -0.408748    	& 14230$\pm$300 & 6.566$\pm$0.105        &0.202$\pm$0.022 & 0.039$\pm$0.005 & 3.6611$\pm$0.1045 &1.8822$\pm$0.3624\\
J0556+0306 	 & 89.157363   & 3.101865 	& 21140$\pm$1510 & 5.388$\pm$0.316       &0.227$\pm$0.028 & 0.159$\pm$0.063 & 0.9427$\pm$0.0669 & 0.9113$\pm$0.6231\\
J0645+1140 	 & 101.383450 & 11.667548 	& 26680$\pm$3110 & 5.385$\pm$0.367      &0.225$\pm$0.030 & 0.159$\pm$0.078 & 0.6939$\pm$0.0648 & 0.4991$\pm$0.4181\\
J0701+0941 	 & 105.262300 & 9.693231   	& 23120$\pm$340 & 5.422$\pm$0.050         &0.225$\pm$0.013 & 0.153$\pm$0.010 & 0.5514$\pm$0.0788 & 0.4378$\pm$0.0408\\
J0737+0259	 & 114.322805 & 2.998566 	& 14580$\pm$450 & 4.667$\pm$0.103       &0.220$\pm$0.018 & 0.360$\pm$0.045 & 0.7804$\pm$0.1644 & 0.3732$\pm$0.0764\\
J0905+3943	 & 136.356200 & 39.727530 	& 15560$\pm$620 & 7.210$\pm$0.261           &0.334$\pm$0.070 & 0.024$\pm$0.008 & 3.3731$\pm$0.1259 & 2.5206$\pm$1.4252\\
J0938+2255	 & 144.599357 & 22.930363 	& 17790$\pm$680 & 6.987$\pm$0.149     &0.293$\pm$0.042 & 0.029$\pm$0.006 & 6.2726$\pm$0.0829 & 2.7432$\pm$0.9480\\
J1129+4715	 & 172.308833 & 47.250481 	& 11410$\pm$70 & 5.181$\pm$0.037        &0.172$\pm$0.011 & 0.178$\pm$0.010 & 1.1961$\pm$0.0392 & 1.0101$\pm$0.0670\\
J1937+3606	 & 294.486970 & 36.107181 	& 25270$\pm$1180 & 5.703$\pm$0.156      &0.209$\pm$0.016 & 0.106$\pm$0.020 & 1.1777$\pm$0.0283   & 1.3599$\pm$0.5121\\
J2116+3746	 & 319.052010 & 37.770494 	& 10360$\pm$140 &5.133$\pm$0.079         &0.164$\pm$0.012   & 0.182$\pm$0.018 & 0.9824$\pm$0.0687   & 0.6121$\pm$0.0753\\
J2257+1216 	 & 344.266849 & 12.279686 	& 18530$\pm$290 & 6.932$\pm$0.084      &0.289$\pm$0.026 & 0.030$\pm$0.003 & 5.8039$\pm$0.1487   & 1.9399$\pm$0.2758\\
J2306+0224 	 & 346.657764 & 2.408254 	& 10810$\pm$70 & 5.126$\pm$0.062          &0.168$\pm$0.011   & 0.185$\pm$0.015 & 1.0712$\pm$0.0987    & 0.6203$\pm$0.0566\\
J2332+5056 	 & 353.233940 & 50.949957 	& 14110$\pm$130 & 4.907$\pm$0.122        &0.201$\pm$0.016   & 0.261$\pm$0.039 & 0.1469$\pm$0.1086    & 0.3702$\pm$0.0722\\
\enddata
\end{deluxetable}

\section{Summary}\label{section5}
We presented the first results from our ongoing project to search for and study ELM WDs using LAMOST low-resolution spectra and $\it Gaia$ data.
This work focused on ELM WD candidates that are included in the catalog of \cite{Pelisoli2019} and the LAMOST DR8 spectral database.  
The major results of this study are as follows: 
\begin{enumerate}
\item A total of 188 LAMOST low-resolution spectra, 172 of which belong to 136 ELM WD candidates and 16 of which come from 12 previously known samples, 
were processed with the spectral fitting program {\sc XTgrid} \citep{Nemeth2012, Nemeth2019}. 
The atmospheric parameters and RV for each of the 188 LAMOST spectra were obtained.
Our spectral fitting results indicate a large variety of stars for the 136 ELM WD candidates, most of which may be hot subdwarf stars or canonical WDs,
and some of which present characteristics of ELM WDs.

\item We compared the atmospheric parameters of the 12 previously known objects from this work to the results published in the ELM Survey paper \citep{Brown2020}.
The atmospheric parameters match particularly well for hot stars with $T_{\rm eff}$ $>$ 15000 K.
Our analysis indicates that the atmospheric fits of objects with $T_{\rm eff}$ $\lesssim$ 15000K are not as reliable, and should be taken just as a proof of concept.
Besides, we have reanalyzed three MMT spectra from the ELM Survey (Warren R. Brown, private communication; \citealt{Brown2020}) to test the validity of our spectral fitting.
The results, derived from two different fits to the same three MMT spectra, are consistent within $1\sigma$ in $\log g$ and $3\sigma$ in $T_{\rm eff}$.

\item Based on the obtained atmospheric parameters and $\it Gaia$ EDR3 data, we found 21 new high-probability ELM WDs 
with masses $M$ $\leq$ 0.3$M_{\sun}$ and parallax estimates that agree to within a factor of 3.
Two of them, J0338+4134 and J1129+4715, show significant RV variability and are very likely to be binary systems containing at least one ELM WD. 
These are the only two high-probability ELM WDs that have more than one spectrum.

\item Our work demonstrates the potential of using the large multi-object spectroscopic facilities of LAMOST to search for and study ELM WDs.
LAMOST RVs from multiple epochs will allow us to confirm their nature as ELM WDs and obtain their physical parameters.
\end{enumerate} 

\begin{acknowledgments}
We thank the anonymous referee for valuable comments. 
This work is supported by the National Natural Science Foundation of China (Grant No. 12003022), the Sichuan Science and Technology Program (Grant No. 2020YFSY0034),
 the Major Science and Technology Project of Qinghai Province (Grant No. 2019-ZJ-A10), 
and the Sichuan Youth Science and Technology Innovation Research Team (Grant No. 21CXTD0038). 
P.N. acknowledges support from the Grant Agency of the Czech Republic (GA\v{C}R 22-34467S).
Y.P.L acknowledges the support from the National Natural Science Foundation of China (Grant No. 12173028), the National Key R\&D Program of China (Grant No. 2021YFA1600401), 
and the Innovation Team Funds of China West Normal University (Grant No. KCXTD2022-6). 
Guoshoujing Telescope (the Large Sky Area Multi-Object Fiber Spectroscopic Telescope LAMOST) is a National Major Scientific Project built by the Chinese Academy of Sciences. 
Funding for the project has been provided by the National Development and Reform Commission. 
LAMOST is operated and managed by the National Astronomical Observatories, Chinese Academy of Sciences. 
This work has made use of data from the European Space Agency (ESA) space mission Gaia. 
Gaia data are being processed by the Gaia Data Processing and Analysis Consortium (DPAC). Funding for the DPAC is provided by national institutions, 
in particular the institutions participating in the Gaia MultiLateral Agreement (MLA).
This research has used the services of \mbox{\url{www.Astroserver.org}} under 
reference IMS1HQ.
\end{acknowledgments}

\software{{\sc XTgrid} \citep{Nemeth2012, Nemeth2019}, TOPCAT\citep{Taylor2005, Taylor2019},
Astropy \citep{Astropy2013, Astropy2018}} 


\end{CJK*}
\end{document}